\documentclass[footinbib,aps,amsmath,amssymb,floatfix,pra,superscriptaddress,longbibliography,showkeys]{revtex4-2}

\usepackage{bm}
\usepackage{ntheorem}
\usepackage{xspace}
\usepackage{threeparttable}
\usepackage{booktabs}

\usepackage[colorlinks]{hyperref}
\hypersetup{%
	plainpages=true,
	breaklinks=true,
	hypertexnames=false,
	pageanchor=true,
	colorlinks=true,
	linkcolor={blue},
	citecolor={blue},
	urlcolor={blue},
	anchorcolor={black}
}
\usepackage{graphicx}   
\usepackage{subfigure}
\usepackage[ruled, linesnumbered]{algorithm2e}
\usepackage{xcolor}
\SetKw{Goto}{goto}

\newcommand{\operatorL}{\mathcal{L}}
\newcommand{\schrodingerisation}{Schr\"{o}dingerisation\xspace}

\newcommand{\operatorN}{\mathcal{N}}
\newcommand{\operatorP}{\mathcal{P}}

\begin{document}



\title{Quantum homotopy analysis method with quantum-compatible linearization for nonlinear partial differential equations}
\author{Cheng Xue}
\affiliation{Institute of Artificial Intelligence, Hefei Comprehensive National Science Center, Hefei, Anhui, 230026, P. R. China}
\author{Xiao-Fan Xu}
\affiliation{CAS Key Laboratory of Quantum Information, University of Science and Technology of China, Hefei, Anhui, 230026, P. R. China}
\author{Xi-Ning Zhuang}
\affiliation{CAS Key Laboratory of Quantum Information, University of Science and Technology of China, Hefei, Anhui, 230026, P. R. China}
\author{Tai-Ping Sun}
\affiliation{CAS Key Laboratory of Quantum Information, University of Science and Technology of China, Hefei, Anhui, 230026, P. R. China}
\author{Yun-Jie Wang}
\affiliation{CAS Key Laboratory of Quantum Information, University of Science and Technology of China, Hefei, Anhui, 230026, P. R. China}
\author{Ming-Yang Tan}
\affiliation{Department of Modern Mechanics, University of Science and Technology of China, Hefei, Anhui, 230026, P. R. China}
\author{Chuang-Chao Ye}
 \affiliation{Origin Quantum Computing Company Limited, Hefei, Anhui, 230026, P. R. China}
 \author{Huan-Yu Liu}
\affiliation{CAS Key Laboratory of Quantum Information, University of Science and Technology of China, Hefei, Anhui, 230026, P. R. China}
\author{Yu-Chun Wu}
 \affiliation{CAS Key Laboratory of Quantum Information, University of Science and Technology of China, Hefei, Anhui, 230026, P. R. China}
\affiliation{Institute of Artificial Intelligence, Hefei Comprehensive National Science Center, Hefei, Anhui, 230026, P. R. China}
\author{Zhao-Yun Chen}
  \email{chenzhaoyun@iai.ustc.edu.cn}
\affiliation{Institute of Artificial Intelligence, Hefei Comprehensive National Science Center, Hefei, Anhui, 230026, P. R. China}
\author{Guo-Ping Guo}
 \affiliation{CAS Key Laboratory of Quantum Information, University of Science and Technology of China, Hefei, Anhui, 230026, P. R. China}
 \affiliation{Institute of Artificial Intelligence, Hefei Comprehensive National Science Center, Hefei, Anhui, 230026, P. R. China}

\begin{abstract}
  
  Nonlinear partial differential equations (PDEs) are crucial for modeling complex fluid dynamics and are foundational to many computational fluid dynamics (CFD) applications. However, solving these nonlinear PDEs is challenging due to the vast computational resources they demand, highlighting the pressing need for more efficient computational methods. Quantum computing offers a promising but technically challenging approach to solving nonlinear PDEs. Recently, Liao proposed a framework that leverages quantum computing to accelerate the solution of nonlinear PDEs based on the homotopy analysis method (HAM), a semi-analytical technique that transforms nonlinear PDEs into a series of linear PDEs. However, the no-cloning theorem in quantum computing poses a major limitation, where directly applying quantum simulation to each HAM step results in exponential complexity growth with the HAM truncation order. This study introduces a ``quantum-compatible linearization'' approach that maps the whole HAM process into a system of linear PDEs, allowing for a one-time solution using established quantum PDE solvers. Our method preserves the exponential speedup of quantum linear PDE solvers while ensuring that computational complexity increases only polynomially with the HAM truncation order. We demonstrate the efficacy of our approach by applying it to the Burgers' equation and the Korteweg-de Vries (KdV) equation. Our approach provides a novel pathway for transforming nonlinear PDEs into linear PDEs, with potential applications to fluid dynamics. This work thus lays the foundation for developing quantum algorithms capable of solving the Navier-Stokes equations, ultimately offering a promising route to accelerate their solutions using quantum computing.

\end{abstract}



\maketitle

\tableofcontents

\section{Introduction}\label{sec1}

Fluid dynamics is an essential discipline for studying the mechanism of flows in engineering and nature. The motion of fluids is typically described by the Navier-Stokes equations (NSEs), which are a set of nonlinear partial differential equations. Due to their nonlinear nature, these equations are generally unsolvable using analytical methods, making them one of the most difficult unsolved problems in mathematics. Due to the difficulty of obtaining an analytical solution, they are usually solved using numerical methods and computers, which has led to the development of computational fluid dynamics. However, the complexity of real-world flows presents significant challenges to both the computational methods used in computational fluid dynamics (CFD) and the performance of computers.

The rapid evolution of computing technology has brought CFD into a new era, especially with recent advancements in GPU-based heterogeneous computing, now enabling large-scale, detailed simulations of certain flows~\cite{witherden2014pyfr}. Nonetheless, according to NASA's 2030 vision for CFD~\cite{slotnick2014cfd}, current computer performance is still inadequate to support larger-scale simulations of practical flows with high-fidelity physics models, such as large eddy simulations (LES) or direct numerical simulations (DNS) of airflow around full-scale aircraft or combustion in jet engines. Additionally, classical computing is approaching physical limits: transistor sizes are nearing the atomic scale~\cite{wu2022vertical}, making it increasingly challenging to design more powerful processors.

To meet growing computational demands, new approaches are essential. Quantum computing, first proposed by Richard Feynman in 1982, operates on the principles of quantum mechanics. With unique properties like superposition and entanglement, quantum computers can tackle specific calculations far more efficiently than classical systems. For instance, Shor’s algorithm~\cite{shor1999polynomial} theoretically offers exponential speedup in breaking RSA encryption—a task that is extraordinarily difficult for classical methods.

As an emerging computational paradigm, quantum computing has shown potential for accelerating the solution of various differential equations, including both partial differential equations (PDEs) and ordinary differential equations (ODEs). Differential equations can be broadly categorized as linear or nonlinear, with quantum computing expected to offer particular advantages for solving linear equations. Approaches include Schrödingerisation~\cite{jin2022quantum,jin2023quantum,jin2024schr}, which maps a linear differential equation to quantum system evolution, algorithms based on quantum linear algebra~\cite{harrow2009quantum,an2022quantum,costa2022optimal}, the linear combination of Hamiltonians~\cite{an2023linear}, analog quantum simulations~\cite{jin2024analog}, and Lindbladian-based methods~\cite{shang2024design}. Quantum solvers for linear PDEs and ODEs have been applied to CFD scenarios, including the hydrodynamic Schrödinger equation~\cite{giannakis2022embedding,meng2023quantum,meng2024simulating}, heat equations~\cite{linden2022quantum}, the Poisson equation~\cite{cao2013quantum,steijl2018parallel,wang2020quantum,liu2021variational}, wave propagation~\cite{chen2024enabling}, and convection-diffusion equations~\cite{budinski2021quantum}.

In contrast, nonlinear differential equations pose a much greater challenge for quantum computers, as quantum computing is inherently linear due to the linear nature of the Schrödinger equation. Despite this challenge, considerable efforts have been made to develop quantum-enhanced approaches for solving nonlinear differential equations.

For nonlinear ODEs, one approach involves the nonlinear transformation of probability amplitudes~\cite{leyton2008quantum}, but this method’s complexity grows exponentially with evolution time, limiting its practicality. To address this, various linearization techniques have been explored. Local linearization combined with intermediate measurements reduces the problem to a sequence of linear problems~\cite{chen2022quantum,joczik2022cost}, although the complexity introduced by intermediate measurements can hinder algorithm performance. The Koopman–von Neumann approach maps nonlinear dynamics to infinite-dimensional linear equations~\cite{joseph2020koopman}, but handling infinite dimensions is practically challenging. Other linearization techniques include Carleman linearization~\cite{liu2021efficient}, coherent state linearization, position-space linearization~\cite{engel2021linear}, and the quantum homotopy perturbation method~\cite{xue2021quantum}. While generally effective for weakly nonlinear equations, these methods may not perform well with strongly nonlinear systems.

For nonlinear PDEs, one strategy is to discretize them into nonlinear ODEs and then apply quantum algorithms developed for ODEs. This includes variational quantum algorithms~\cite{lubasch2020variational,kyriienko2021solving,sarma2024quantum,huang2023near} and techniques utilizing quantum amplitude estimation~\cite{oz2023efficient,gaitan2021finding}. Another approach is to directly linearize nonlinear PDEs into linear PDEs, such as using the level set method to transform Hamilton–Jacobi and scalar hyperbolic PDEs~\cite{jin2024quantum}. However, this approach is limited to specific types of PDEs and cannot be readily generalized.
The main challenges in this field include the limited applicability of current linearization techniques to nonlinear PDEs, as most methods are designed for ODEs. Extending these techniques to a broader class of nonlinear PDEs will require further research. Additionally, existing linearization methods are generally suitable only for weakly nonlinear equations, so developing quantum algorithms capable of efficiently solving strongly nonlinear equations remains a significant hurdle.

Recently, Liao proposed a quantum framework for solving nonlinear PDEs~\cite{liao2024general} based on the homotopy analysis method (HAM)~\cite{liao1992proposed,liao2003beyond,liao2004homotopy}, a semi-analytical approach capable of handling a wide range of PDEs, including those with strong nonlinearity. However, this initial work has yet to be fully developed into a quantum algorithm, with many implementation details remaining open. In this paper, we present a comprehensive analysis of this promising approach and propose a quantum algorithm for implementing the homotopy analysis method on a quantum computer. Specifically, we introduce a “quantum-compatible linearization” technique that maps the entire HAM process into a system of linear PDEs, enabling a single solution using established quantum PDE solvers. Building upon this quantum-compatible linearization, we develop a Quantum Homotopy Analysis Method (QHAM). The QHAM preserves the exponential speedup offered by quantum linear PDE solvers while ensuring that computational complexity increases only polynomially with the HAM truncation order. We validate our approach by applying it to the Burgers' equation and the Korteweg–de Vries (KdV) equation.

The paper is organized as follows. In Section~\ref{sec:qham-algorithm}, we review the concept of HAM and conduct a preliminary analysis to identify key challenges for quantum implementation. We then introduce our proposed quantum homotopy analysis method (QHAM), presenting a novel technique to bridge the gap between HAM and quantum computation. Section~\ref{sec:qham-complexity} provides a detailed analysis of the computational complexity of the proposed method. To improve convergence, we propose a quantum-iteration approach for iterative execution of HAM in Section~\ref{sec:qham-iterative}. In Section~\ref{sec:application}, we demonstrate the capability of QHAM through applications to the Burgers' and KdV equations. Finally, as a more challenging example, we address the Navier-Stokes (NS) equation in Section~\ref{sec:qham-NS}, outlining a potential pathway for quantum computation of the NS equation and its remaining challenges. Conclusions and further discussions are presented in Section~\ref{sec:conclusion}.

\section{Quantum homotopy analysis method}\label{sec:qham-algorithm}

In this paper, we aim to implement a quantum version of the homotopy analysis method (HAM), enabling the solution of nonlinear PDEs using a quantum computer. To begin, we provide an overview of HAM~\cite{liao1992proposed, liao2003beyond, liao2004homotopy,liao2020new}. Developed by Liao in the 1990s, HAM extends traditional perturbation methods by introducing a homotopy—a continuous deformation—between a solvable problem and the original nonlinear problem. This approach allows for flexible solutions without relying on small parameters, making it particularly valuable for solving nonlinear problems where conventional perturbation techniques fail.

\begin{figure}[h]
    \centering
    \includegraphics[width=16.9cm]{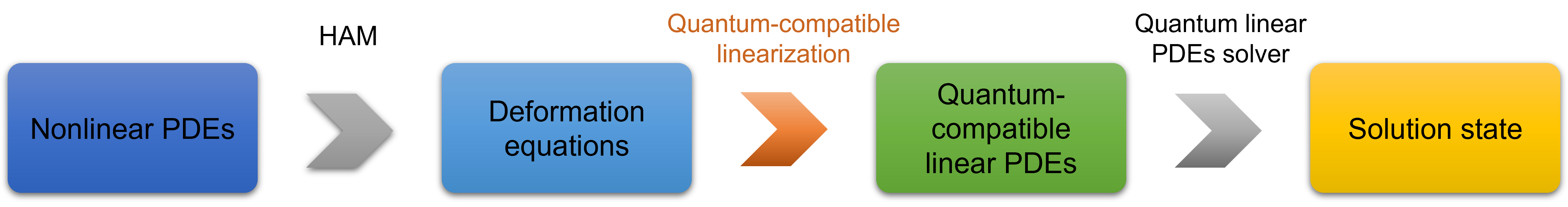}
    \caption{QHAM Framework. The framework consists of three main steps: HAM, quantum-compatible linearization, and a quantum linear PDE solver. The first two steps transform the nonlinear PDEs into quantum-compatible linear PDEs. In the final step, quantum algorithms are employed to accelerate the solution of the quantum-compatible linear PDEs, with solutions to the nonlinear PDEs obtained through post-selection.}
    \label{fig:qham-framework}
\end{figure}

To outline our proposed algorithm, we present the main workflow here, with detailed descriptions provided in the following sections. Our algorithm consists of three primary steps, as illustrated in Figure~\ref{fig:qham-framework}. First, a standard HAM is applied to transform the nonlinear PDE problem into a series of deformation equations for \( U_i \). Next, we analyze these deformation equations and highlight the challenges they present for quantum computation. To address these challenges, we introduce a novel approach called ``quantum-compatible linearization", which embeds the deformation equations for \( U_i \) into quantum-compatible linear PDEs. At this stage, the linearized PDEs can be solved using established quantum algorithms, such as Schrödingerisation, to obtain the output state.

\subsection{Homotopy analysis method}

HAM begins by transforming a nonlinear PDE into a series of deformation equations. Given a nonlinear PDE expressed as
\begin{equation}
    \frac{\partial u}{\partial t} = \operatorN(u),
\end{equation}
where $\operatorN$ denotes a nonlinear operator, $u=u(x,t)$ is the variable, and subject to the initial condition
\begin{equation}
    u(x,t=0)=u_{\mathrm{in}}.
\end{equation}
HAM constructs a homotopy in the form
\begin{equation}\label{eq-homotopy}
    \mathcal{H}(U, q) = (1-q)\operatorL(U - U_0) - q h H(t) \left(\frac{\partial U}{\partial t}-\operatorN(U)\right) = 0,
\end{equation}
where \( q \in [0, 1] \) is an embedding parameter, \( \operatorL \) is a linear operator satisfying \( \operatorL(0) = 0 \), $U_0$ is the initial guess for the solution, $h$ is the convergence-control parameter, and $H(t)$ is an adjustable auxiliary function. The function $U$ satisfies the conditions
\begin{equation}
    U(x, t;q=0)=U_0, \quad U(x, t;q=1)=u.
\end{equation}
In HAM, $\operatorL$, $h$, $U_0$, and $H(t)$ are all adjustable, offering flexibility in tailoring the method to achieve convergence and accuracy in solving nonlinear PDEs.

The solution \(U\) can be expanded by a series in terms of the embedding parameter \(q\): 
\begin{equation}
    U(t, x; q) = U_0(x, t) + \sum_{i=1}^{\infty} q^i U_i(x, t),
\end{equation}
where $U_i(x,t)$ represents the $i$th-order basis function. When $q=1$, we recover the original solution as 
\begin{equation}\label{eq-ham-series}
    u(x, t) = U_0(x, t) + \sum_{i=1}^{\infty} U_i(x, t).
\end{equation}
By selecting appropriate HAM parameters to ensure the convergence of the series~\cite{liao2003beyond}, each $U_i(x,t)$ satisfies the following condition
\begin{equation}
    \Vert U_{i+1}(x, t) \Vert \leq \alpha \Vert U_i(x, t) \Vert, \quad \alpha < 1
\end{equation}
for any evolution time $t$. Here \( \alpha \) is a parameter controlling the convergence rate, \( \Vert U_{i+1}(x, t) \Vert \) denotes the $L^2$ norm, which is defined as 
\begin{equation}
    \Vert U_{i+1}(x, t) \Vert_{L^2}=(\int_{x}{U_{i+1}^2(x, t)dx})^{1/2}.
\end{equation}
In this paper, when $f$ is a function, $\Vert f \Vert=\Vert f \Vert_{L^2}$, and when $f$ is a vector, $\Vert f \Vert=\Vert f \Vert_{2}=(\sum_{i}{f_i^2})^{1/2}$. 

To approximate the solution, we truncate the series expansion at order \( m \), yielding:
\begin{equation}\label{eq-utq}
    U(x, t; q) \approx U_0(x, t) + \sum_{i=1}^{m} q^i U_i(x, t).
\end{equation}
The choice of truncation order $m$ influences the accuracy of the solution. Specifically, the error introduced by truncation is given by
\begin{equation}
    \left\Vert \sum_{j=m+1}^{\infty} U_j(x, t) \right\Vert \leq \frac{\alpha^{m+1}}{1 - \alpha} \left\Vert U_0(x, t) \right\Vert.
\end{equation}
We define \( \beta =\max_{t}{\Vert U_0(x, t) \Vert} \). Given an error \( \epsilon \), the truncation order is chosen as follows:
\begin{equation}\label{eq-m-bound}
    m = \left\lceil \log_{1/\alpha}{\frac{\beta}{(1 - \alpha)\epsilon}} - 1 \right\rceil.
\end{equation}

The task now involves solving for \( U_0 \) through \( U_m \) (a total of \( m + 1 \) functions), with the final solution \( U \) obtained as the sum of these functions. By substituting Eq.~\eqref{eq-utq} into Eq.~\eqref{eq-homotopy} and differentiating with respect to \( q \) up to order \( m \), we set \( q = 0 \) to derive the \( i \)th-order deformation equation:
\begin{equation}\label{eq:deformation}
    \operatorL(U_i - \chi_i U_{i-1}) = h H(t) R_i(t),
\end{equation}
where \( \chi_1 = 0 \), \( \chi_i = 1 \) for \( i > 1 \), and \( R_i(t) \) is defined as
\begin{equation}\label{eq-ri}
    R_i(t) = \left. \frac{1}{(i-1)!} \frac{d^{i-1} (\partial_t U-\operatorN(U))}{dq^{i-1}} \right|_{q=0}.
\end{equation}

This formulation enables us to rewrite Eq.~(\ref{eq:deformation}) in a recursive form:
\begin{equation}\label{eq-t-deformation}
    \left\{
    \begin{aligned}
        \operatorL(U_1) &= h H(t) R_1(t), \\[6pt]
        \operatorL(U_2) &= h H(t) [R_1(t) + R_2(t)], \\[6pt]
                        & \vdots \\[6pt]
        \operatorL(U_m) &= h H(t) \sum_{i=1}^{m} R_i(t).
    \end{aligned}\right.
\end{equation}
This system of equations systematically constructs each \( U_i \) term, leading to the solution \( U \) as the sum of these iterative components.

\subsection{Quantum-compatible linearization}\label{sec-second-linearization}

Now we analyze the properties of Eq.~\eqref{eq-t-deformation} through the lens of quantum computing. Notably, \( R_i \) contains only terms \( U_0, U_1, \ldots, U_{i-1} \), allowing Eq.~\eqref{eq-t-deformation} to be treated as a linear PDE. Each \( U_i \) can be solved using a quantum linear PDE solver, producing the quantum state \( |U_i\rangle \). However, due to the quantum no-cloning theorem, \( |U_i\rangle \) cannot be reused multiple times. During the solution process for \( U_i \), the solver requires multiple queries to \( |U_k\rangle \) (\( k < i \)), meaning that \( |U_k\rangle \) (\( k < i \)) must be prepared repeatedly. Consequently, the overall complexity grows exponentially with \( m \). Therefore, further linearization of the deformation equations is crucial for developing an efficient quantum simulation approach.

To address the issue discussed above, we introduce an additional transformation to eliminate nonlinear terms and reduce the computational cost associated with iterative solutions. In this section, we propose a method called ``quantum-compatible linearization'', which embeds the \( m+1 \) deformation equations into linear PDEs that can be efficiently processed on a quantum computer.

To establish a unified approach to nonlinear PDEs, we start by examining quadratic nonlinear PDEs with first-order time derivatives. Quadratic nonlinear PDEs are a specific class where the highest-order terms consist of quadratic combinations of derivatives. For a variable \( u(x,t) \), this quadratic nonlinear PDE can be represented as:
\begin{equation}
    \operatorN(u) := \operatorN_0 + \operatorN_1(u) + \operatorN_2(u).
\end{equation}
In this formulation, \( \operatorN_i \) denotes the \( i \)th-order polynomial operator involving \( u \) or its derivatives. The three types of operators are defined as follows: \( \operatorN_0 \), which serves as the driving term; \( \operatorN_1 \), a linear operator; and \( \operatorN_2 \), a nonlinear operator. The nonlinear term \( \operatorN_2(U) \) can be further decomposed as:
\begin{equation}\label{eq-n2}
    \operatorN_2(U) = \sum_{k=1}^{s} \operatorL_{1,k}(U) \operatorL_{2,k}(U),
\end{equation}
where \( \operatorL_{1,k} \) and \( \operatorL_{2,k} \) are homogeneous linear operators, and \( s \) represents the number of terms in this decomposition.

To apply HAM to quadratic nonlinear PDEs, we select \( \operatorL \) and \( U_0 \) as follows:
\begin{equation}\label{eq-u0}
    \frac{\partial U_0}{\partial t} = \operatorN_0 + \operatorN_1(U_0), \quad U_0(0) = u_{\text{in}},
\end{equation}
and
\begin{equation}\label{eq-lu}
    \operatorL(U) = \partial_t U - \operatorN_1(U).
\end{equation}
Here, \( U_0 \) serves as the initial guess solution, evolving according to the driving term \( \operatorN_0 \) and the linear operator \( \operatorN_1 \). The choice of \( \operatorL \) and \( U_0 \) is flexible, enabling adjustments that ensure convergence and accuracy within HAM.

Our quantum-compatible linearization method begins by analyzing the nonlinear components in the deformation equations. For the \( i \)th-order deformation equation, the nonlinear components are embedded in \( R_i(t) \). To proceed, we derive the expression for \( R_i(t) \). According to Eq.~(\ref{eq-ri}), for the quadratic nonlinear PDE within HAM, \( R_i(t) \) is written as
\begin{equation}\label{eq-rit}
    R_i(t) = \left. \frac{1}{(i-1)!} \frac{d^{i-1} (\partial_t U - \operatorN_0 - \operatorN_1(U) - \operatorN_2(U))}{dq^{i-1}} \right|_{q=0}.
\end{equation}
We have
\begin{equation}
    \left. \frac{1}{(i-1)!} \frac{d^{i-1} U}{dq^{i-1}} \right|_{q=0} = U_{i-1}, \quad i > 0.
\end{equation}
Consequently,
\begin{equation}\label{eq-u-n1}
    \left.\frac{1}{(i-1)!} \frac{d^{i-1} (\partial_t U - \operatorN_1(U))}{dq^{i-1}}\right|_{q=0} = \partial_t U_{i-1} - \operatorN_1(U_{i-1}), \quad i > 0.
\end{equation}
By substituting Eq.~(\ref{eq-utq}) into Eq.~(\ref{eq-n2}), we expand \( \operatorN_2(U) \) as follows:
\begin{equation}\label{eq-n2u}
    \operatorN_2(U) = \sum_{j=0}^{m} \sum_{l=0}^{m} \sum_{k=1}^{s} \operatorL_{1,k}(U_j)\operatorL_{2,k}(U_l) q^{j+l}.
\end{equation}
Utilizing Theorem 2.2 from \cite{liao2009notes}, we derive
\begin{equation}\label{eq-n2-exp}
    \left.\frac{1}{(i-1)!} \frac{d^{i-1} \operatorN_2(U)}{dq^{i-1}}\right|_{q=0} = \sum_{j=0}^{i-1}\sum_{k=1}^{s} \operatorL_{1,k}(U_j)\operatorL_{2,k}(U_{i-1-j}), \quad i > 0.
\end{equation}
Combining Eqs.~(\ref{eq-t-deformation}), (\ref{eq-u0}), (\ref{eq-lu}), (\ref{eq-u-n1}), (\ref{eq-n2-exp}), and (\ref{eq-rit}), we obtain the recursive relations for \( R_i(t) \):

For \( i = 1 \):
\begin{equation}\label{eq:deformation_R1}
\begin{aligned}
    R_1(t) &= \partial_t U_0 - \left( \operatorN_0 + \operatorN_1(U_0) + \operatorN_2(U_0) \right) \\
           &= -\operatorN_2(U_0) \\
           &= -\sum_{k=1}^{s} \operatorL_{1,k}(U_0)\operatorL_{2,k}(U_0).
\end{aligned}
\end{equation}

For \( i > 1 \):
\begin{equation}\label{eq:deformation_Ri}
\begin{aligned}
    R_i(t) &= \partial_t U_{i-1} - \operatorN_1(U_{i-1}) - \left.\frac{1}{(i-1)!} \frac{d^{i-1} \operatorN_2(U)}{dq^{i-1}}\right|_{q=0} \\
           &= \operatorL(U_{i-1}) - \sum_{j=0}^{i-1} \sum_{k=1}^{s} \operatorL_{1,k}(U_j)\operatorL_{2,k}(U_{i-1-j})\\
           &=h H(t) \sum_{l=1}^{i-1} R_l(t) - \sum_{j=0}^{i-1} \sum_{k=1}^{s} \operatorL_{1,k}(U_j)\operatorL_{2,k}(U_{i-1-j}).
\end{aligned}
\end{equation}
This recursive relation can be explicitly solved and expressed as
\begin{equation}\label{eq:deformation_R_v1}
    R_i(t) = -\sum_{j=0}^{i-1} \sum_{k=1}^{s} \operatorL_{1,k}(U_j)\operatorL_{2,k}(U_{i-1-j}) - h H(t)\sum_{l=0}^{i-2}(1+hH(t))^{i-2-l}\sum_{j=0}^{l} \sum_{k=1}^{s} \operatorL_{1,k}(U_j)\operatorL_{2,k}(U_{l-j}), \quad i\geq 1.
\end{equation}

Thus, the nonlinear components in the \( i \)th-order deformation equations are expressed as \( \operatorL_{1,k}(U_j)\operatorL_{2,k}(U_{l-j}) \) for \( l=0,1,\cdots,i-1,\quad j = 0, 1, \cdots, l \).

In the quantum-compatible linearization process, we introduce new variables to transform the nonlinear components \( \operatorL_{1,k}(U_j)\operatorL_{2,k}(U_{l-j}) \) into linear functions of these newly defined variables. This is accomplished through dimension expansion: since \( U_j \) and \( U_{l-j} \) are defined in \( x \)-space, denoted here as \( x_0 \), we introduce an auxiliary space \( x_1 \). We then define a new variable \( W = U_j(x_0)U_{l-j}(x_1) \), which satisfies
\begin{equation}\label{eq-sec-eq1}
    \operatorL_{1,k}(U_j(x_0))\operatorL_{2,k}(U_{l-j}(x_0)) = \delta_{x_0,x_1} \operatorL_{1,k,x_0}(\operatorL_{2,k,x_1}(W)),
\end{equation}
where $\operatorL_{i,k,x_j}$ means that the $\operatorL_{i,k}$ operator acts on $x_j$ space, $\delta_{x_0,x_1}$ is defined as
\begin{equation}
    \delta_{x_0,x_1}=\left\{
        \begin{array}{cc}
             1,& x_0=x_1,\\
             0,& else.
        \end{array}   
    \right.
\end{equation}
Equation~(\ref{eq-sec-eq1}) demonstrates that the term \( \operatorL_{1,k}(U_j(x_0))\operatorL_{2,k}(U_{l-j}(x_0)) \) in the \( i \)th-order deformation equations is linear with respect to \( W \). To proceed, we derive the PDEs for the new variable \( W \), expressed as  
\begin{equation}\label{eq-new-w}
    (\operatorL_{x_0,t} + \operatorL_{x_1,t})(W) = \operatorL_{x_0,t}(U_j(x_0)) U_{l-j}(x_1) + U_j(x_0) \operatorL_{x_1,t}(U_{l-j}(x_1)),
\end{equation}
where \( \operatorL_{x_i,t} \) denotes \( \operatorL \) in the \((x_i, t)\) space. Substituting Eq.~(\ref{eq-t-deformation}) into the right side of Eq.~(\ref{eq-new-w}) generates new nonlinear components, such as \( U_{l}(x_0) U_{j-1-l}(x_0) U_{l-j}(x_1) \). We can linearize these generated nonlinear components by constructing additional new variables in a similar manner. 

The right side of Eq.~(\ref{eq-new-w}) also contains \( \operatorN_0 \), such as $\operatorN_0(x_0)U_{l-j}(x_1)$. The variable $\operatorN_0(x_0)U_{l-j}(x_1)$ is not a new variable, it can be converted to $U_{l-j}(x_0)\operatorN_0(x_1)$ with a permutation operator \( \operatorP \), which is defined as
\begin{equation}
\operatorP_{0,2,3,\cdots,i,1}\left(U_{a_0}(x_0) \operatorN_0(x_i) U_{a_2}(x_1) \cdots U_{a_i}(x_{l})\right) = U_{a_0}(x) \operatorN_0(x_1) \cdots U_{a_i}(x_i).
\end{equation}
Here, \( \operatorP_{b_0, b_1, \cdots, b_i} \) denotes permuting \( (x_0, x_1, \cdots, x_i) \) to \( (x_{b_0}, x_{b_1}, \cdots, x_{b_i}) \).
Because $U_{l-j}(x_0)$ appears in the quantum-compatible linearization of the $(l-j)$th-order deformation equations and $\operatorN_0(x_1)$ is known, $U_{l-j}(x_0)\operatorN_0(x_1)$ is not a new variables. 
Since \( \operatorP \) is a linear operator, it is compatible with the quantum-compatible linearization process, ensuring that the non-homogeneous term \( \operatorN_0 \) does not introduce new variables.

\begin{figure}
    \centering
    \includegraphics[width=16.9cm]{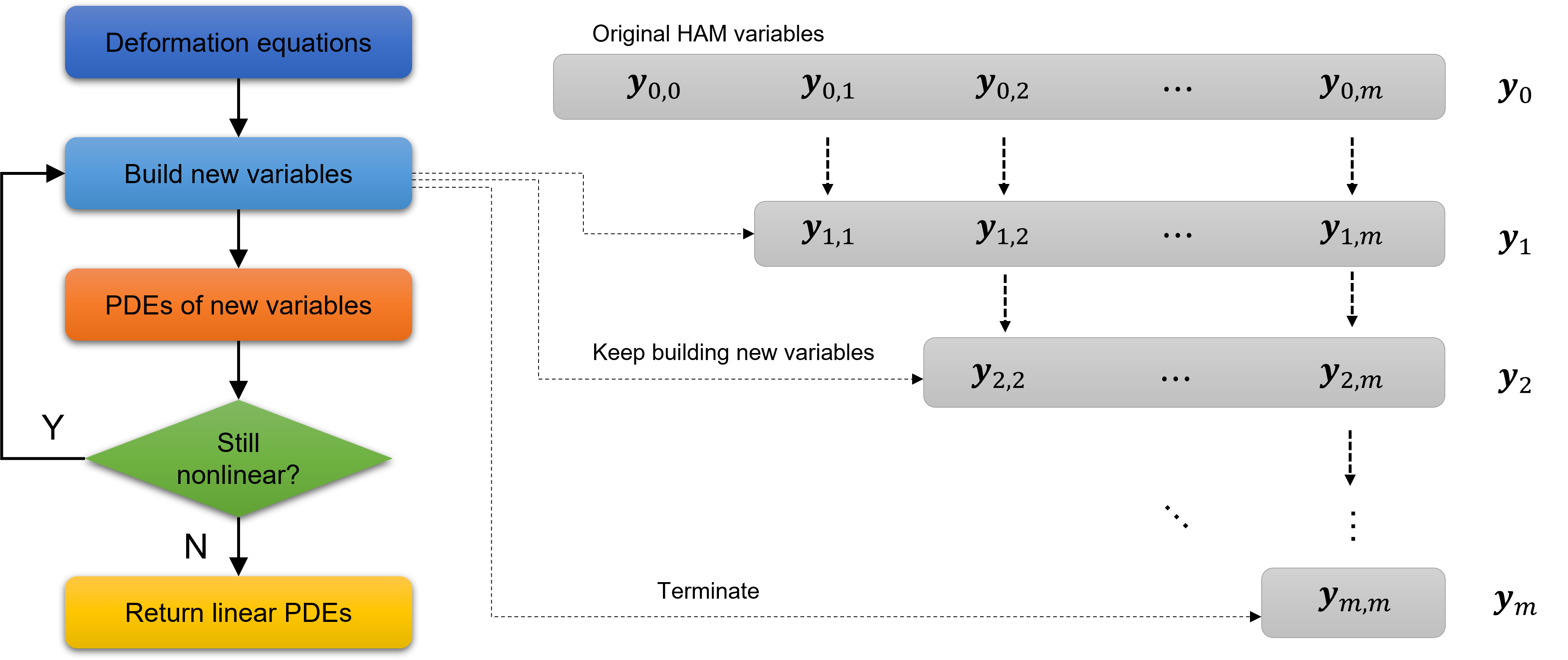}
    \caption{Quantum-compatible linearization of deformation equations. $\boldsymbol{y}_{0,i}=U_i$ are the original HAM variables. The quantum-compatible linearization process is applied to $i$th-order deformation equations from $i=1$ to $i=m$. The quantum-compatible linearization of the $i$th-order deformation equations will terminate at the variable $\boldsymbol{y}_{i,i}=\Pi_{j=0}^{i}{U_0(x_j)}$.}
    \label{fig:secondary-linearization}
\end{figure}

The quantum-compatible linearization repeats this process: constructing new variables for each nonlinear term and deriving the associated equations. If the resulting equations still contain nonlinear terms, we continue constructing additional variables based on these terms. This iterative process continues until all nonlinear terms are eliminated, as shown on the left side of Figure~\ref{fig:secondary-linearization}.

A key question in the quantum-compatible linearization process is whether it can fully eliminate all nonlinear terms. Consider an intermediate variable of the form \( \prod_{j=0}^{k} U_{a_j}(x_j) \) that arises during the process. This variable introduces new terms of the form \( \prod_{j=0}^{k+1} U_{b_j}(x_j) \), governed by the relationship 
\begin{equation}
    \sum_{j=0}^{k} a_j = 1 + \sum_{j=0}^{k+1} b_j.
\end{equation}
Consequently, for \( m \)th-order deformation equations, the quantum-compatible linearization process will ultimately yield the variable \( \Pi_{j=0}^{m+1} U_0(x_j) \). Since \( U_0 \) satisfies linear PDEs, \( \Pi_{j=0}^{m+1} U_0(x_j) \) will not generate any additional variables, ensuring that the quantum-compatible linearization process comes to a complete halt. This termination guarantees that all nonlinear terms are effectively eliminated, making the resulting system amenable to efficient quantum computation.

\begin{algorithm}[H]\label{ham-seondary-linearization}
\caption{HAM with quantum-compatible linearization}
Use HAM to construct deformation equations up to \( m \)th-order\;
Define $\boldsymbol{y}_{0,0}=U_0$\;
\For{$j \gets 1$ \KwTo $m$}{
    Define $i \gets 0$, $\boldsymbol{y}_{i,j} \gets U_j$, \;
    \eIf{Check the $\boldsymbol{y}$-related equation ($\boldsymbol{y}$ contains all constructed $\boldsymbol{y}_{i,j}$) is linearized is \textbf{true}}{
        $\bm{\text{Break}}$
        
    }{
        Increment order: \( i \gets i + 1 \)\;
        Build variable \( \boldsymbol{y}_{i,j} \)\;
        Derive \( \boldsymbol{y}_{i,j} \)-related PDEs\;
        \Goto~Step 5\;
    } 
}
Return $\boldsymbol{y}$-related linear PDEs\;
\end{algorithm}

As introduced in Algorithm~\ref{ham-seondary-linearization}, by applying the quantum-compatible linearization process to the \( i \)-th deformation equations for \( i = 1 \) to \( i = m \), we construct a system of linear PDEs encompassing the variables \( U_0, U_1, \ldots, U_m \) and all newly introduced variables. We define the complete set of variables as \( \boldsymbol{y} \), whose structure is illustrated in Figure~\ref{fig:secondary-linearization}. The variable \( \boldsymbol{y} \) is represented as
\begin{equation}
    \boldsymbol{y} = [\boldsymbol{y}_0, \boldsymbol{y}_1, \boldsymbol{y}_2, \ldots, \boldsymbol{y}_m],
\end{equation}
with
\begin{equation}
    \boldsymbol{y}_0 = [\boldsymbol{y}_{0,0}, \boldsymbol{y}_{0,1}, \ldots, \boldsymbol{y}_{0,m}], \quad \boldsymbol{y}_{0,j} = U_j(x_0), \quad j = 0, 1, \ldots, m.
\end{equation}

Here, \( \boldsymbol{y}_{ij} \) represents the newly constructed variables arising from equations related to \( \boldsymbol{y}_{i-1,j} \). We find that the variables in \( \boldsymbol{y}_{ij} \) satisfy
\begin{equation}\label{eq-yij}
    \left\{U_{a_0}(x_0) U_{a_1}(x_1) \cdots U_{a_i}(x_i) \mid a_k \geq 0, \sum_{k=0}^{i} a_k = j - i \right\},
\end{equation}
where \( x_1, x_2, \ldots, x_i \) represent auxiliary (ancilla) spaces. From Eq.~(\ref{eq-yij}), we can determine that the number of variables in \( \boldsymbol{y}_{i,j} \) is \( \binom{j}{i} \).

Notably,
\begin{equation}
    \boldsymbol{y}_{i,i} = U_0(x_0) U_0(x_1) \cdots U_0(x_i),
\end{equation}
and since the equation involving \( U_0 \) is a linear PDE, the equations associated with \( \boldsymbol{y}_{i,i} \) do not contain nonlinear components. Thus, the quantum-compatible linearization process terminates when \( i = j \), ensuring that no further nonlinear terms are introduced.
A more specific example demonstrating the quantum-compatible linearization process will be provided in Section~\ref{sec:application}.

Finally, we return to the original PDEs and establish the relationship between the transformed equation and the original nonlinear PDEs. By substituting \( q = 1 \) into Eq.~(\ref{eq-utq}), we find that the solution to the original nonlinear PDEs is \( u = \sum_{i=0}^{m} U_i \). We can define \( u \) as a new variable \( \boldsymbol{y}_{-1} \) and extend \( \boldsymbol{y} \) as follows:
\begin{equation}
    \boldsymbol{y} = [\boldsymbol{y}_{-1}, \boldsymbol{y}_{0}, \boldsymbol{y}_{1}, \cdots, \boldsymbol{y}_{m}].
\end{equation}

Since 
\begin{equation}
    \boldsymbol{y}_{-1} = \sum_{i=0}^{m} \boldsymbol{y}_{i,0}, 
\end{equation}
the expanded \( \boldsymbol{y} \) still satisfies a system of linear PDEs. We express the \( \boldsymbol{y} \)-related PDEs as
\begin{equation}\label{eq-linearPDE}
    \partial_t \boldsymbol{y} = \operatorL^{*}(\boldsymbol{y}) + \boldsymbol{b}(t), \quad \boldsymbol{y}(0) = \boldsymbol{y}_{\text{in}},
\end{equation}
the details of the linear operator \( \operatorL^{*} \), the inhomogeneous terms $\boldsymbol{b}(t)$, and the initial condition $\boldsymbol{y}_{\text{in}}$ are derived from the quantum-compatible linearization process. In specific, $\boldsymbol{b}(t)$ and $\boldsymbol{y}_{\text{in}}$ are written as 
\begin{equation}\label{eq-bt}
    \boldsymbol{b}_{-1}(t)=\operatorN_0, \boldsymbol{b}_{0,0}(t)=\operatorN_0, \boldsymbol{b}_{i,j}(t)=0, j=1,\cdots,m, i=0,1,\cdots,j.
\end{equation}
\begin{equation}\label{eq-yin}
    \boldsymbol{y}_{-1,\text{in}}=u_{\text{in}}(x_0), \boldsymbol{y}_{i,i,\text{in}}=\Pi_{j=0}^{i}{u_{\text{in}}(x_i)},i=0,1,2,\cdots,m.
\end{equation}

\subsubsection{Quantum-compatible linearization for HAM truncation order $m=1$}
To illustrate the implementation process of the quantum-compatible linearization technique, we consider the case when HAM truncation order \( m = 1 \). 

For \( i = 0 \), we define \( \boldsymbol{y}_{0,0} = U_0(x_0) \), which satisfies the equation
\begin{equation}\label{eq-y00}
    \partial_t \boldsymbol{y}_{0,0} = \operatorN_{0,x_0} + \operatorN_{1,x_0}(\boldsymbol{y}_{0,0}), \quad \boldsymbol{y}_{0,0,\text{in}} = u_{\text{in}}(x_0),
\end{equation}
where \( \operatorN_{j,x_k} \) denotes the operator \( \operatorN_j \) acting on the spatial variable \( x_k \).

For \( i = 1 \), we have \( \boldsymbol{y}_{0,1} = U_1(x_0) \), which satisfies
\begin{equation}\label{eq-1st-deformation-equations}
    \partial_t \boldsymbol{y}_{0,1} = \operatorN_{1,x_0}(\boldsymbol{y}_{0,1}) - h H(t) \sum_{k=1}^{s} \operatorL_{1,k,x_0}(U_0(x_0)) \operatorL_{2,k,x_0}(U_0(x_0)).
\end{equation}
Since Eq.~(\ref{eq-1st-deformation-equations}) contains nonlinear terms of the form \( \operatorL_{1,k,x_0}(U_0(x_0)) \operatorL_{2,k,x_0}(U_0(x_0)) \), we introduce a new variable \( \boldsymbol{y}_{1,1} = U_0(x_0) U_0(x_1) \) to facilitate linearization.

Using this new variable, Eq.~(\ref{eq-1st-deformation-equations}) can be rewritten as
\begin{equation}\label{eq-y01}
    \partial_t \boldsymbol{y}_{0,1} = \operatorN_{1,x_0}(\boldsymbol{y}_{0,1}) - h H(t) (\sum_{k=1}^{s} \delta_{x_0,x_1} \operatorL_{1,k,x_0} \operatorL_{2,k,x_1}) (\boldsymbol{y}_{1,1}).
\end{equation}

Next, we derive an equation for \( \boldsymbol{y}_{1,1} \). Starting from \( \boldsymbol{y}_{1,1} = U_0(x_0) U_0(x_1) \), we compute its time derivative:
\begin{align}\label{eq-y11-derivation}
    \partial_t \boldsymbol{y}_{1,1} &= \partial_t \left( U_0(x_0) U_0(x_1) \right) \notag \\
    &= \left( \partial_t U_0(x_0) \right) U_0(x_1) + U_0(x_0) \left( \partial_t U_0(x_1) \right).
\end{align}
Using the governing equations for \( U_0(x_0) \) and \( U_0(x_1) \):
\begin{equation}\label{eq-U0-eq}
    \partial_t U_0(x_k) = \operatorN_{0,x_k} + \operatorN_{1,x_k}(U_0(x_k)), \quad k = 0,1,
\end{equation}
we substitute into Eq.~(\ref{eq-y11-derivation}) to obtain
\begin{align}
    \partial_t \boldsymbol{y}_{1,1} &= \left( \operatorN_{0,x_0} + \operatorN_{1,x_0}(U_0(x_0)) \right) U_0(x_1) + U_0(x_0) \left( \operatorN_{0,x_1} + \operatorN_{1,x_1}(U_0(x_1)) \right) \notag \\
    &= \operatorN_{0,x_0} U_0(x_1) + U_0(x_0) \operatorN_{0,x_1} + \operatorN_{1,x_0}(U_0(x_0)) U_0(x_1) + U_0(x_0) \operatorN_{1,x_1}(U_0(x_1)).
\end{align}
Recognizing that \( \operatorN_{1,x_0}(U_0(x_0)) U_0(x_1) = \operatorN_{1,x_0}(\boldsymbol{y}_{1,1}) \) and similarly for \( \operatorN_{1,x_1} \), we simplify to
\begin{align}
    \partial_t \boldsymbol{y}_{1,1} &= \left( \operatorN_{1,x_0} + \operatorN_{1,x_1} \right) (\boldsymbol{y}_{1,1}) + \operatorN_{0,x_0} U_0(x_1) + U_0(x_0) \operatorN_{0,x_1}.
\end{align}
Using the symmetry between \( x_0 \) and \( x_1 \), we can write
\begin{equation}
    \operatorN_{0,x_0} U_0(x_1) + U_0(x_0) \operatorN_{0,x_1} = \left( 1 + \operatorP_{x_0,x_1} \right) \left( U_0(x_0) \operatorN_{0,x_1} \right),
\end{equation}
where \( \operatorP_{x_0,x_1} \) is the permutation operator exchanging \( x_0 \) and \( x_1 \).
Therefore, the equation for \( \boldsymbol{y}_{1,1} \) becomes
\begin{equation}\label{eq-y11}
    \partial_t \boldsymbol{y}_{1,1} = \left( \operatorN_{1,x_0} + \operatorN_{1,x_1} \right) (\boldsymbol{y}_{1,1}) + \left( 1 + \operatorP_{x_0,x_1} \right) \left( \boldsymbol{y}_{0,0} \operatorN_{0,x_1} \right).
\end{equation}
This equation contains only linear terms involving \( \boldsymbol{y}_{1,1} \) and \( \boldsymbol{y}_{0,0} \), completing the linearization of the first-order deformation equations.

We now define \( \boldsymbol{y}_{-1} = \boldsymbol{y}_{0,0} + \boldsymbol{y}_{0,1} \), which satisfies
\begin{align}\label{eq-y-1}
    \partial_t \boldsymbol{y}_{-1} &= \partial_t \boldsymbol{y}_{0,0} + \partial_t \boldsymbol{y}_{0,1} \notag \\
    &= \operatorN_{0,x_0} + \operatorN_{1,x_0}(\boldsymbol{y}_{0,0}) + \operatorN_{1,x_0}(\boldsymbol{y}_{0,1}) - h H(t) (\sum_{k=1}^{s} \delta_{x_0,x_1} \operatorL_{1,k,x_0} \operatorL_{2,k,x_1})( \boldsymbol{y}_{1,1}) \notag \\
    &= \operatorN_{0,x_0} + \operatorN_{1,x_0}(\boldsymbol{y}_{-1}) - h H(t) (\sum_{k=1}^{s} \delta_{x_0,x_1} \operatorL_{1,k,x_0} \operatorL_{2,k,x_1})( \boldsymbol{y}_{1,1}).
\end{align}
The initial condition is \( \boldsymbol{y}_{-1,\text{in}} = \boldsymbol{y}_{0,0,\text{in}} + \boldsymbol{y}_{0,1,\text{in}} = u_{\text{in}}(x_0) + 0 = u_{\text{in}}(x_0) \).

Collecting Eqs.~(\ref{eq-y00}), (\ref{eq-y01}), (\ref{eq-y11}), and (\ref{eq-y-1}), we obtain the system of linear partial differential equations:
\begin{equation}\label{eq-linear-PDEs}
    \left\{
    \begin{aligned}
        \partial_t \boldsymbol{y}_{-1} &= \operatorN_{0,x_0} + \operatorN_{1,x_0}(\boldsymbol{y}_{-1}) - h H(t) (\sum_{k=1}^{s} \delta_{x_0,x_1} \operatorL_{1,k,x_0} \operatorL_{2,k,x_1} )(\boldsymbol{y}_{1,1}), & \boldsymbol{y}_{-1,\text{in}} &= u_{\text{in}}(x_0), \\
        \partial_t \boldsymbol{y}_{0,0} &= \operatorN_{0,x_0} + \operatorN_{1,x_0}(\boldsymbol{y}_{0,0}), & \boldsymbol{y}_{0,0,\text{in}} &= u_{\text{in}}(x_0), \\
        \partial_t \boldsymbol{y}_{0,1} &= \operatorN_{1,x_0}(\boldsymbol{y}_{0,1}) - h H(t) (\sum_{k=1}^{s} \delta_{x_0,x_1} \operatorL_{1,k,x_0} \operatorL_{2,k,x_1})( \boldsymbol{y}_{1,1}), & \boldsymbol{y}_{0,1,\text{in}} &= 0, \\
        \partial_t \boldsymbol{y}_{1,1} &= \left( \operatorN_{1,x_0} + \operatorN_{1,x_1} \right) \boldsymbol{y}_{1,1} + \left( 1 + \operatorP_{x_0,x_1} \right) \left( \boldsymbol{y}_{0,0} \operatorN_{0,x_1} \right), & \boldsymbol{y}_{1,1,\text{in}} &= u_{\text{in}}(x_0) u_{\text{in}}(x_1).
    \end{aligned}
    \right.
\end{equation}
This system represents the quantum-compatible linearization of the first-order deformation equations. All nonlinear terms have been successfully expressed in terms of linear operators acting on the introduced variables \( \boldsymbol{y}_{-1} \), \( \boldsymbol{y}_{0,0} \), \( \boldsymbol{y}_{0,1} \), and \( \boldsymbol{y}_{1,1} \).

\subsubsection{Structure analysis of the quantum-compatible linear PDEs}

We now analyze the structure of Eq.~(\ref{eq-linearPDE}). The variable \( \boldsymbol{y}_{-1} \) consists of a single component, while each \( \boldsymbol{y}_{i,j} \) contains \( \binom{j}{i} \) variables, corresponding to the number of ways to choose \( i \) elements from \( j \). Therefore, the total number of variables in \( \boldsymbol{y} \) is given by
\begin{equation}\label{eq-total-variables}
    1 + \sum_{i=0}^{m} \sum_{j=i}^{m} \binom{j}{i} = 2^{m+1}.
\end{equation}
This implies that \( \operatorL^{*} \) can be represented as a \( 2^{m+1} \times 2^{m+1} \) matrix.

Next, we analyze the sparsity of \( \operatorL^{*} \), which is characterized by the maximum number of non-zero entries in any row of the matrix. This corresponds to the maximum number of variables in \( \boldsymbol{y} \) that are directly related to any given variable through the linear operator \( \operatorL^{*} \).

Consider a variable in \( \boldsymbol{y}_{i,j} \) with \( i \geq 0 \), which can be expressed in the form $\prod_{k=0}^{i} U_{a_k}(x_k)$,
where the indices \( a_k \) satisfy the condition \( \sum_{k=0}^{i} a_k \leq m - i \). 
From Eq.~(\ref{eq-t-deformation}) and \( \operatorL_{x_k,t} = \partial_t - \operatorN_{1,x_k} \), we have 
\begin{align}\label{eq-yij-time-derivative}
    \partial_t \left( \prod_{k=0}^{i} U_{a_k}(x_k) \right)&=\sum_{k=0}^{i}  U_{a_0}(x_0) \dots [\partial_{t} U_{a_k}(x_k)] \dots U_{a_i}(x_i)  \notag\\
    &= \left( \sum_{k=0}^{i} \operatorN_{1,x_k} \right) \left( \prod_{k=0}^{i} U_{a_k}(x_k) \right)  + \sum_{k=0}^{i} \left( U_{a_0}(x_0) \dots \left[ \operatorL_{x_k,t}(U_{a_k}(x_k)) \right] \dots U_{a_i}(x_i) \right)\notag\\
    &= \left( \sum_{k=0}^{i} \operatorN_{1,x_k} \right) \left( \prod_{k=0}^{i} U_{a_k}(x_k) \right)  + \sum_{k=0}^{i} \left( U_{a_0}(x_0) \dots \left[ h H(t) \sum_{l=1}^{a_k} R_l(t,x_k) \right] \dots U_{a_i}(x_i) \right).
\end{align}

The terms involving \( R_l(t,x_k) \) generate new variables. Specifically, each term \( U_{a_0}(x_0) \dots \left[ h H(t) R_l(t,x_k) \right] \dots U_{a_i}(x_i) \) produces new variables of the form
\begin{equation}\label{eq-new-variables}
    U_{a_0}(x_0) \dots U_{b}(x_k) U_{l - b}(x_{i+1}) \dots U_{a_i}(x_i), \quad \text{for } l = 0, 1, \dots, a_k - 1, \quad b = 0, 1, \dots, l,
\end{equation}
where \( x_{i+1} \) denotes a newly introduced ancilla space. From Eq.~(\ref{eq-new-variables}), we derive that the total number of new variables generated from \( U_{a_k}(x_k) \) is \( \frac{a_k(a_k + 1)}{2} \). 
Therefore, the number of variables directly related to \( U_{a_0}(x_0) U_{a_1}(x_1) \dots U_{a_i}(x_i) \) is $\sum_{k=0}^{i}{a_k(a_k + 1)/2}$.
Since \( \sum_{k=0}^{i} a_k \leq m - i \), this sum is maximized when one \( a_k \) takes the value \( m - i \) and the others are zero, giving an upper bound of $\frac{(m - i)(m - i + 1)}{2}$. 

In the matrix representation of \( \operatorL^{*} \), the variable \( U_{a_0}(x_0) U_{a_1}(x_1) \dots U_{a_i}(x_i) \) corresponds to a specific row, while each new variable \( U_{a_0}(x_0) \dots U_{b}(x_k) U_{l - b}(x_{i+1}) \dots U_{a_i}(x_i) \) corresponds to a column. The corresponding non-zero element of \( \operatorL^{*} \) at this row and column position is given by
\begin{equation}\label{eq-Lstar-element}
    -h H(t) \delta_{x_k, x_{i+1}} \sum_{j=1}^{s} \operatorL_{1,j,x_k} \operatorL_{2,j,x_{i+1}},
\end{equation}

Since \( \boldsymbol{y}_{-1} = \sum_{i=0}^{m} \boldsymbol{y}_{i,0} \), the variables related to \( \boldsymbol{y}_{-1} \) correspond to the new variables generated by \( R_i(t) \) for \( i = 0, 1, \dots, m \). The total number of such variables is $\sum_{i=1}^{m} i =m(m+1)/2$.

Based on the above analysis, we can determine the precise positions and expressions of each non-zero element in \( \operatorL^{*} \).

\subsection{Incorporating QHAM with quantum linear PDEs solvers}

After applying quantum-compatible linearization, we obtain the \( \boldsymbol{y} \)-related linear PDEs, which can be solved using quantum algorithms. Numerous quantum algorithms have been developed for solving linear PDEs, commonly referred to as quantum linear PDE solvers. The initial step in most quantum linear PDE solvers involves discretizing the linear PDEs into linear ODEs. 

\renewcommand{\arraystretch}{2}
\begin{table}[htbp]
    \centering    
    \caption{Query complexity of quantum linear ODEs solver \cite{an2023quantum,shang2024design}. Here $\alpha_A\geq \Vert A(t)\Vert$, $T$ is the evolution time, $\epsilon$ is the error, and $\beta\in (0,1)$. $\kappa_V$ is the upper bound of condition numbers of $A(T)$, $\Delta$ is defined as the lower bound of the smallest non-zero eigenvalues of the Hermitian part of $A(T)$.}
    \begin{threeparttable}
        \begin{tabular}{c|c}
            \toprule
            \toprule
            Algorithm & Query complexity to $A(t)$\\
            \hline
            Quantum spectral methods~\cite{childs2020quantum} &  $\widetilde{\mathcal{O}}\left(\frac{\Vert Y_{\text{in}}\Vert}{\Vert Y(T)\Vert}\alpha_A \kappa_A T \text{polylog}(1/\epsilon)\right)$\\
            \hline
            Truncated Dyson series \cite{berry2024quantum} & $\widetilde{\mathcal{O}}\left(\frac{\Vert Y_{\text{in}}\Vert}{\Vert Y(T)\Vert} \alpha_A T (\log(1/\epsilon))^2\right)$\\
            \hline
            \schrodingerisation~\cite{jin2212quantum} & $\widetilde{\mathcal{O}}\left(\frac{\Vert Y_{\text{in}} \Vert}{\Vert Y(T)\Vert} \alpha_A T/\epsilon\right)$\\
            \hline
            Time-marching \cite{berry2024quantum} & $\widetilde{\mathcal{O}}\left(\frac{\Vert Y_{\text{in}}\Vert}{\Vert Y(T)\Vert} \alpha^2_A T^2 \log(1/\epsilon)\right)$\\
            \hline
            Original LCHS \cite{an2023linear} & $\widetilde{\mathcal{O}}\left(\left(\frac{\Vert Y_{\text{in}}\Vert}{\Vert Y(T)\Vert}\right)^2 \alpha_A T/\epsilon\right)$\\
            \hline
            Improved LCHS \cite{an2023quantum} & $\widetilde{\mathcal{O}}\left(\frac{\Vert Y_{\text{in}} \Vert}{\Vert Y(T)\Vert} \alpha_A T (\log(1/\epsilon))^{1+1/\beta}\right)$\\
            \hline
            Lindbladians \cite{shang2024design} & $\widetilde{\mathcal{O}}\left(\frac{\Vert Y_{\text{in}} \Vert}{\Vert Y(T)\Vert} \alpha_A \Delta^{-1}T \frac{\log^{3}{1/\epsilon}}{\log^2{\log(1/\epsilon)}}\right)$\\
            \bottomrule
            \bottomrule
        \end{tabular}
        \label{tab:complexity_linear}
    \end{threeparttable}
\end{table}
\renewcommand{\arraystretch}{1}

We employ the finite difference method~\cite{thomas1995numerical} to discretize Eq.~(\ref{eq-linearPDE}). Specifically, each subspace \( x_i \) for \( i = 0, 1, \dots, m \) is discretized into \( n \) grid points, denoted as \( [x_{i,0}, x_{i,1}, \dots, x_{i,n-1}] \). The discretized linear ODEs are then written as
\begin{equation}\label{eq-linear-ODEs}
    \frac{dY}{dt} = A(t) Y + B(t), \quad Y(0) = Y_{\text{in}},
\end{equation}
where \( Y = [Y_{-1}, Y_0, \dots, Y_m] \), and \( Y_i \), \( B(t) \), and \( Y_{\text{in}} \) represent the discretized vectors of \( \boldsymbol{y}_i \), \( \boldsymbol{b}(t) \), and \( \boldsymbol{y}_{\text{in}} \), respectively. The matrix \( A(t) \) is the discretized form of the operator \( \operatorL^{*} \).

We analyze the dimension of Eq.~(\ref{eq-linear-ODEs}), denoted as \( N \). Note that the dimension of \( Y_{-1} \) is \( n \). As introduced in previous sections, the number of variables in \( \boldsymbol{y}_{i,j} \) is \( \binom{j}{i} \) for \( i \geq 0 \), and the dimension of the discretized vector of a variable in \( \boldsymbol{y}_{i,j} \) is \( n^{i+1} \). In summary, the dimension \( N \) is given by
\begin{equation}
    N = n + \sum_{i=0}^{m} \sum_{j=i}^{m} \binom{j}{i} n^{i+1} = (n+1)^{m+1} + n - 1.
\end{equation}

Next, we solve Eq.~(\ref{eq-linear-ODEs}) using quantum linear ODE solvers. Table~\ref{tab:complexity_linear} summarizes the complexities of several typical quantum linear ODE solvers proposed in recent years. To utilize these solvers, we need to construct the input oracles for Eq.~(\ref{eq-linear-ODEs}). For any \( t \in [0, T] \), the following oracles should be constructed:
\begin{equation}
\begin{aligned}\label{eq-oracles}
    O_{A,1}:& \ |i\rangle|j\rangle \mapsto |i\rangle|F_{i,j}\rangle, \quad i, j \in [0, N-1],\\
    O_{A,2}:& \ |i\rangle|j\rangle|0\rangle \mapsto |i\rangle|j\rangle|A_{i,j}\rangle, \quad i, j \in [0, N-1],\\
    O_B:& \ |0\rangle \mapsto \frac{1}{\Vert B(t) \Vert} \sum_{i=0}^{N-1} B(t)_i |i\rangle,\\
    O_{\text{in}}:& \ |0\rangle \mapsto \frac{1}{\Vert Y_{\text{in}} \Vert} \sum_{i=0}^{N-1} Y_{\text{in},i} |i\rangle,    
\end{aligned}
\end{equation}
where \( F_{i,j} \) represents the column index of the \( j \)-th non-zero element in the \( i \)-th row of \( A(t) \). Some quantum linear ODE solvers require additional input oracles. For example, the truncated Dyson series solver~\cite{berry2024quantum} requires a block-encoding~\cite{gilyen2019quantum} of \( A(t) \), which can be constructed by querying \( O_{A,1} \) and \( O_{A,2} \) a constant number of times~\cite{chakraborty2018power,gilyen2019quantum}. Other required oracles can also be constructed using the oracles defined in Eq.~(\ref{eq-oracles}). Therefore, we only need to construct the oracles specified in Eq.~(\ref{eq-oracles}).

The construction processes of these oracles are as follows:
\begin{itemize}
    \item \( O_{A,1} \), \( O_{A,2} \): The matrix \( A(t) \) is the discretization of \( \operatorL^{*} \). The sparsity of \( A(t) \) is related to the sparsity of \( \operatorL^{*} \), and the discretization process affects the sparsity as well. Specifically, the sparsity increases linearly with the number of terms in each element of \( \operatorL^{*} \). The number of off-diagonal elements in \( \operatorL^{*} \) is at most \( s \), where \( s \) is the number of terms in \( \operatorN_2 \). The order of the finite difference method also affects the sparsity, but since it is generally a constant, we ignore its influence here. Therefore, the sparsity of \( A(t) \) is the sparsity of \( \operatorL^{*} \) multiplied by \( s \), that is, \( s m(m+1)/2 \), indicating that \( A(t) \) remains a sparse matrix. Furthermore, each element of \( A(t) \) can be obtained in \( \mathcal{O}(s m^2) \) time, and the complexity to construct \( O_{A,1} \) and \( O_{A,2} \) is \( \mathcal{O}(s m^2) \). This process is similar to the oracle construction in \cite{xue2021quantum}.
    \item \( O_B \): The vector \( B(t) \) is the discretization of \( \boldsymbol{b}(t) \), which is a \( 2^{m+1} \)-dimensional vector as described in Eq.~(\ref{eq-bt}). Given the oracle \( O_{\operatorN_0} \) that prepares the amplitude-encoded state of the discretized \( \operatorN_0 \), \( O_B \) can be constructed by querying \( O_{\operatorN_0} \) a constant number of times. Generally, when \( \operatorN_0 \) is an integrable function, the discretized \( \operatorN_0 \) can be prepared in \( \mathcal{O}(\text{polylog}(n)) \) time~\cite{grover2002creating}.
    \item \( O_{\text{in}} \): The vector \( Y_{\text{in}} \) is the discretization of \( \boldsymbol{y}_{\text{in}} \), a \( 2^{m+1} \)-dimensional vector as given in Eq.~(\ref{eq-yin}). Given the oracle \( O_{y0} \) that prepares the amplitude-encoded state of the discretized \( \boldsymbol{y}_{\text{in}} \), \( O_{\text{in}} \) can be constructed by querying \( O_{y0} \) \( \mathcal{O}(m) \) times. Similarly, when \( \boldsymbol{y}_{\text{in}} \) is an integrable function, the discretized \( \boldsymbol{y}_{\text{in}} \) can be prepared in \( \mathcal{O}(\text{polylog}(n)) \) time~\cite{grover2002creating}.
\end{itemize}

We then input the above oracles into a high-performance solver, such as the truncated Dyson series solver. For this solver, the query complexity of \( A(t) \) is given by
\begin{equation}
    \widetilde{\mathcal{O}}\left( \frac{\Vert Y_{\text{in}} \Vert}{\Vert Y(T) \Vert} \alpha_A T \log^2(1/\epsilon) \right),
\end{equation}
where \( \alpha_A \) is a parameter related to the norm of \( A(t) \), \( T \) is the evolution time, and \( \epsilon \) is the error tolerance. In quantum linear ODE solvers, the query complexities of \( O_B \) and \( O_{\text{in}} \) are less than that of \( O_{A,1} \) and \( O_{A,2} \), so they do not significantly affect the overall complexity expression and can be ignored.

Upon executing the quantum linear ODE solver, we obtain the output state
\begin{equation}
    |Y(T)\rangle = \frac{1}{\Vert Y(T) \Vert} \sum_{i=-1}^{m} |i\rangle \otimes \left( \Vert Y_i(T) \Vert |Y_i(T)\rangle \right),
\end{equation}
where
\begin{align}
    |Y_{-1}(T)\rangle &= \frac{1}{\Vert Y_{-1}(T) \Vert} \sum_{j=0}^{n-1} Y_{-1,j} |j\rangle, \notag\\
    |Y_i(T)\rangle &= \frac{1}{\Vert Y_i(T) \Vert } \sum_{j=i}^{m} \sum_{k=0}^{\binom{j}{i}} \sum_{l=0}^{n^{i+1}} Y_{i,j,k,l} |i, j, k, l\rangle, \quad i = 0, 1, \dots, m,
\end{align}
and \( Y_{-1} \) is the \( n \)-dimensional discretized vector of \( \boldsymbol{y}_{-1} \), \( Y_{i,j,k} \) is the \( n^{i+1} \)-dimensional discretized vector of the \( k \)-th variable in \( \boldsymbol{y}_{i,j} \), and \( Y_{i,j,k,l} \) is the \( l \)-th element in \( Y_{i,j,k} \).

Since \( \boldsymbol{y}_{-1}(T) \) is the HAM solution, i.e., \( \boldsymbol{y}_{-1}(T) = u(T) \), \( |Y_{-1}(T)\rangle \) represents the quantum state of the original nonlinear PDE solution. To isolate the solution state \( |Y_{-1}(T)\rangle \), we measure the first register of \( |Y(T)\rangle \) in the \( |-1\rangle \) state, which collapses the second register to \( |u(T)\rangle \). This measurement step is probabilistic. In the next section, we will analyze the success rate and overall complexity of the QHAM.

\section{Resource analysis}\label{sec:qham-complexity}

\subsection{Time complexity}

The contribution of the QHAM's time complexity consists of three parts.
\begin{itemize}
    \item $T_A$: \textbf{The complexity of implementing $A(t)$.}
    \item $T_Q$: \textbf{Query complexity of $A(t)$ for quantum linear PDE solver.}
    \item $p$: \textbf{Success rate to post select $|Y_{-1}(T)\rangle$}.
\end{itemize}

The final complexity is thus given by:
\begin{equation}
    T_A\cdot T_Q \cdot \frac{1}{\sqrt{p}}.
\end{equation}

As introduced in the previous section, the complexity of implementing $A(t)$ is 
\begin{equation}
    T_A=\mathcal{O}(sm^2),
\end{equation}
and one of the optimal $T_Q$ is 
\begin{equation}
    T_Q=\widetilde{\mathcal{O}}\left( \frac{\Vert Y_{\text{in}}\Vert}{\Vert Y(T)\Vert} \alpha_A T \log^2(1/\epsilon)\right).
\end{equation}

Here we first analyze the success rate to post select $|Y_{-1}(T)\rangle$.
Denoted as $p$, the success rate is computed as 
\begin{equation}
    p=\Vert Y_{-1}(T)\Vert^2/\Vert Y(T) \Vert^2,
\end{equation}
using amplitude amplification, the success rate is amplified to $\Omega(1)$ with $\mathcal{O}(1/\sqrt{p})=\mathcal{O}(\Vert Y(T)\Vert^2/\Vert Y_{-1}(T) \Vert^2)$ query complexity. 
Since $Y$ is constructed from $U_i$, and to ensure the convergence of the homotopy analysis, the norm of $U_i$ should decrease with increasing $i$. We ensure this by selecting appropriate homotopy analysis parameters, then $U_i$ satisfies
\begin{equation}\label{eq-ham-convergence}
    \Vert U_{i+1} \Vert \leq \Vert U_i \Vert \alpha, \quad \alpha < 1.
\end{equation}
From $Y_{-1}=\sum_{i=0}^{m}{U_i}$, we have
\begin{equation}
    \left(1 - \frac{\alpha(1 - \alpha^m)}{1 - \alpha}\right) \Vert U_0 \Vert \leq \Vert Y_{-1} \Vert \leq \frac{1 - \alpha^{m+1}}{1 - \alpha} \Vert U_0 \Vert,
\end{equation}
and to ensure $\Vert Y_{-1}\Vert > 0$, $\alpha$ must satisfy $\alpha < 1/2$. Furthermore, it can be shown:
\begin{equation}
    \Vert Y_{ij}\Vert \leq \binom{j}{i} \alpha^{j-i} \Vert U_0 \Vert^{i+1},
\end{equation}
where
\begin{equation}
\begin{aligned}
    \Vert Y \Vert &\leq \Vert Y_{-1}\Vert + \sum_{i=0}^{m}{\sum_{j=i}^{m}{\Vert Y_{ij}\Vert }}\\[8pt]
    &\leq \Vert Y_{-1}\Vert + \sum_{i=0}^{m}{\sum_{j=i}^{m}{\binom{j}{i} \alpha^{j-i} \Vert U_0 \Vert^{i+1} }}\\[8pt]
    &\leq \frac{1 - \alpha^{m+1}}{1 - \alpha} \Vert U_0 \Vert + \frac{1 - (\alpha + \Vert U_0 \Vert)^{m+1}}{1 - (\alpha + \Vert U_0 \Vert)} \Vert U_0 \Vert,
\end{aligned}
\end{equation}
To prevent $\Vert Y \Vert$ from diverging, we require $\alpha < 1 - \Vert U_0 \Vert$. Therefore, we can infer:
\begin{equation}
     \sqrt{p}=\frac{\Vert Y_{-1}(T)\Vert}{\Vert Y(T) \Vert}\geq \left(\frac{1 - 2\alpha}{1 - \alpha}\right) \bigg/ \left(\frac{1}{1 - \alpha} + \frac{1}{1 - (\alpha + \| U_0 \|)}\right) = \frac{(1 - 2\alpha)(1 - \alpha - \| U_0 \|)}{2 - 2\alpha - \| U_0 \|}.
\end{equation}
$\sqrt{p}$ is influenced by $\alpha$ and $\Vert U_0 \Vert$. $\alpha$ represents the convergent factor of the QHAM, by selecting appropriate HAM parameters, such as initial guess solution, convergence factor, etc., the HAM convergence speed can be improved, thereby reducing $\alpha$. $\Vert U_0 \Vert$ can be adjusted by scaling the original nonlinear PDEs. For example, we define $v=\eta u$ and obtain the $v$-related nonlinear PDEs, then $\Vert V_0 \Vert=\eta\Vert U_0 \Vert$, $\eta$ can be adjusted to make $\Vert V_0 \Vert$ small enough and has little effect on $\sqrt{p}$.


Therefore, the query complexity of the QHAM is 
\begin{equation}
    T_Q/\sqrt{p}=\widetilde{\mathcal{O}}\left(\frac{2 - 2\alpha - \| U_0(T) \|}{(1 - 2\alpha)(1 - \alpha - \| U_0(T) \|)} \frac{\Vert Y_{\text{in}}\Vert}{\Vert Y(T)\Vert} \alpha_A T \log^2(1/\epsilon) \right).
\end{equation}
The gate complexity is the query complexity multiplied by a factor $sm^2 \text{polylog}(N)$. As introduced before, 
\begin{equation}
    m= \left\lceil \log_{1/\alpha}{\frac{\beta}{(1 - \alpha)\epsilon}} - 1 \right\rceil, N=(n+1)^{m+1}+n- 1,
\end{equation}
so that we have
\begin{equation}
    m^2 \text{polylog}(N) \sim \mathcal{O}(\text{polylog}(n/\epsilon)).
\end{equation}
Thus the gate complexity becomes
\begin{equation}\label{eq-time-complexity}
    \widetilde{\mathcal{O}}\left(\frac{2 - 2\alpha - \| U_0(T) \|}{(1 - 2\alpha)(1 - \alpha - \| U_0(T) \|)} \frac{s\Vert Y_{\text{in}}\Vert\alpha_A }{\Vert Y(T)\Vert}  \text{polylog}(n/\epsilon) \right).
\end{equation}

The time complexity of the classical HAM is $\tilde{\mathcal{O}}(mnT\log(1/\epsilon))$. Although QHAM provides exponential acceleration on $n$, its dependence on $\epsilon$ is worse than the classical HAM. Worse, the QHAM is also influenced by other parameters, such as $\alpha$, $\alpha_A$, $\frac{\Vert Y_{\text{in}}\Vert}{\Vert Y(T)\Vert}$. $\alpha$ can be adjusted by choosing suitable HAM parameters, the factor $\frac{\Vert Y_{\text{in}}\Vert}{\Vert Y(T)\Vert}\alpha_A$ is derived from $\frac{\max_{t\in [0,T]}{\Vert Y(t)\Vert}}{\Vert Y(T)\Vert}$, when the solution $Y(t)$ is relatively stable, the factor $\frac{\Vert Y_{\text{in}}\Vert}{\Vert Y(T)\Vert}\alpha_A$ has little effect on the complexity of the QHAM. 

\subsection{Qubit number}

Now we analyze the qubit number in the QHAM. The dimension of the linear ODEs defined in Eq.~(\ref{eq-linear-ODEs}) is $N\approx (n+1)^{m+1}$. The size of the space during the execution of quantum linear ODEs solver is $TN$, and the required qubit number to represent the space is $\mathcal{O}(\log(TN))=\mathcal{O}(\log(T)+m\log(n))$. Furthermore, the ancilla qubits are required to finish intermediate operations, such as quantum arithmetic, the number of the ancilla qubits is $\mathcal{O}(\log(1/\epsilon))$.

In summary, the total qubit number is $\mathcal{O}(\log(T/\epsilon)+m\log(n))$.

\section{Iterative QHAM}\label{sec:qham-iterative}

The choice of the initial guess solution \( U_0 \) influences the convergence of the HAM. When \( U_0 \) is unsuitable, the QHAM might not converge; we refer to this as the "bad initial guess solution" problem. One way to select a suitable \( U_0 \) is to execute the HAM iteratively. Specifically, we use the solution obtained from the HAM as the new initial guess and then execute the HAM again to obtain an updated solution. We repeat this step until we achieve convergence.

In our method, we construct an Iterative QHAM (IQHAM) to linearize the iteration process. The linearized PDEs of the first iteration are shown in Eq.~(\ref{eq-linearPDE}), where \( \boldsymbol{y}_{-1} = U \) is the HAM solution. We then set \( \boldsymbol{y}_{-1} \) as the initial guess solution \( \widetilde{U}_0 \) and execute the HAM again, where \( \widetilde{U}_{i} \) represents the solution of the \( i \)-th deformation equation. Next, we linearize the \( \widetilde{U}_{i} \)-related deformation equations using the quantum-compatible linearization strategy and obtain the following PDEs:
\begin{equation}\label{eq-2iter-lpde-1}
    \partial_t \widetilde{\boldsymbol{y}} = \operatorL^{*}(\widetilde{\boldsymbol{y}}) + \widetilde{\boldsymbol{b}}(t),
\end{equation}
where the structure of \( \widetilde{\boldsymbol{y}} \) is the same as that of \( \boldsymbol{y} \), except that the variable \( U_i \) is replaced by \( \widetilde{U}_i \). Note that \( \widetilde{\boldsymbol{y}} \) contains variables related to \( \widetilde{U}_0 = \boldsymbol{y}_{-1} \). We can directly replace \( \widetilde{U}_0 \) in \( \widetilde{\boldsymbol{y}} \) with \( \boldsymbol{y} \), except in \( \widetilde{\boldsymbol{y}}_{-1} \). For example, a term \( \widetilde{U}_3(x_0) \widetilde{U}_0(x_1) \) changes to \( \widetilde{U}_3(x_0) \boldsymbol{y}(x_1, x_2, \dots, x_{m+1}) \). We define the modified \( \widetilde{\boldsymbol{y}} \) as \( \boldsymbol{z} \); the \( \boldsymbol{z} \)-related PDEs remain linear and can be derived by combining Eqs.~(\ref{eq-linearPDE}) and (\ref{eq-2iter-lpde-1}). The variable \( \boldsymbol{z} \) contains \( \boldsymbol{z}_{-1} = \sum_{i=0}^{m} \widetilde{U}_i \), which represents the target solution of the original nonlinear PDEs obtained from the iterative QHAM. Therefore, the \( \boldsymbol{z} \)-related PDEs are solved using quantum linear PDE solvers, and the quantum state of the target solution is obtained after the post-selection process.

The solution of the \( \boldsymbol{z} \)-related PDEs can be regarded as the new initial guess solution, and the HAM is performed again. We can linearize the entire process in the same way. Consequently, we linearize the process of iteratively executing QHAM and obtain a system of linear PDEs. Finally, we solve these linear PDEs with the quantum linear PDE solver.

The above describes the IQHAM implementation method. Now we analyze the performance of the IQHAM. We consider the iteration number \( l \), with \( \boldsymbol{z}^{(l)} \) representing the variables of the linearized PDEs. When \( l = 0 \), \( \boldsymbol{z}^{(0)} = \boldsymbol{y} \) corresponds to the original QHAM; \( l = 1 \) means the process has been iterated once, and so on.

As introduced earlier, when \( l = 0 \), the number of variables in \( \boldsymbol{z}^{(0)} \) is \( 2^{m+1} \). When \( l = 1 \), the number of variables becomes \( \mathcal{O}(2^{(m+1)^2}) \). Consequently, the number of variables in the \( l \)-iteration IQHAM is \( \mathcal{O}(2^{(m+1)^l}) \).

Next, we analyze the success rate of the IQHAM. Considering only the main components, we derive that \( \boldsymbol{z}^{(l)} \) satisfies
\begin{equation}
    \Vert \boldsymbol{z}^{(l)} \Vert \sim \frac{1 - (\alpha + \Vert \boldsymbol{z}^{(l-1)} \Vert)^{m+1}}{1 - (\alpha + \Vert \boldsymbol{z}^{(l-1)} \Vert)} \Vert \boldsymbol{z}^{(l-1)} \Vert,
\end{equation}
where \( \boldsymbol{z}^{(-1)} = U_0 \). However, as \( l \) increases, the upper bound of \( \Vert \boldsymbol{z}^{(l)} \Vert \) exceeds 1, causing the success rate to decrease exponentially with \( m \).

Therefore, the space and time complexity of the IQHAM increase exponentially with \( l \). The IQHAM is a trade-off to mitigate the problem of a "bad" initial guess solution in QHAM. To make IQHAM effective, the iteration number should be as small as possible. Fortunately, in specific problems, the iteration number can be kept small. In Section~\ref{sec-kdv}, we use the IQHAM to solve the KdV equations and provide numerical evidence.

\section{Applications}\label{sec:application}

\subsection{Burgers' equation}

\subsubsection{Formulation}

The Burgers' equation is a fundamental partial differential equation used to model various physical processes in fluid mechanics, nonlinear acoustics, and traffic flow. It combines nonlinear convection and diffusion terms, making it valuable for studying shock waves, turbulence, and other nonlinear phenomena.

Here, we consider the one-dimensional forced Burgers' equation:
\begin{equation}
    u_t + u u_x = \mu u_{xx} + f(x), \quad 0 \leq x \leq 1, \quad 0 \leq t \leq 1,
\end{equation}
where \( u = u(x, t) \) represents the velocity field, $f(x)=0.3 \cos(\pi x)$, and \( \mu = 0.1 \) is the viscosity coefficient characterizing the strength of viscous effects. The term \( 0.3 \cos(\pi x) \) is an external forcing function.

The initial and boundary conditions are determined by the exact solution:
\begin{equation}
    u(x, 0) = 0.3 \sin(\pi x), \quad u(0, t) = 0, \quad u(1, t) = 0.
\end{equation}
To solve this equation using the Quantum Homotopy Analysis Method (QHAM), we first construct a homotopy:
\begin{equation}
    \mathcal{H}(q, U) = (1 - q) \operatorL(U - U_0) - h H(t) q \operatorN(U) = 0,
\end{equation}

We choose the initial guess \( U_0(x, t) \) to satisfy:
\begin{equation}
    (U_0)_t = \mu (U_0)_{xx} + f(x),
\end{equation}
with initial and boundary conditions:
\begin{equation}
    U_0(x, 0) = 0.3 \sin(\pi x), \quad U_0(0, t) = 0, \quad U_0(1, t) = 0.
\end{equation}
The solution \( U(x, t; q) \) is expanded as a power series in \( q \):
\begin{equation}
    U(x, t; q) = U_0(x, t) + \sum_{i=1}^{m} q^i U_i(x, t).
\end{equation}
At \( q = 1 \), this series converges to the solution of the original equation:
\begin{equation}
    u(x, t) = U(x, t; q = 1) = U_0(x, t) + \sum_{i=1}^{m} U_i(x, t).
\end{equation}
The \( m \)-th order deformation equation is given by:
\begin{equation}
    \operatorL(U_m - \chi_m U_{m-1}) = h H(t) R_m(t),
\end{equation}
where
\begin{align}
    &R_1(t) = (U_0)_t - \mu (U_0)_{xx} + U_0(U_0)_x - f = U_0(U_0)_x, \\
    &R_m(t) = (U_{m-1})_t - \mu (U_{m-1})_{xx} + \sum_{i=0}^{m-1} U_i (U_{m-1-i})_x, \quad m > 1.
\end{align}

Next, we set the truncation order \( m = 1 \) and apply the quantum-compatible linearization process. We introduce the variables:

\begin{align}
    \boldsymbol{y}_{-1} &= U_0(x_0) + U_1(x_0), \notag\\[6pt]
    \boldsymbol{y}_0 &= \begin{bmatrix} U_0(x_0) \\ U_1(x_0) \end{bmatrix}, \\[6pt]
    \boldsymbol{y}_1 &= U_0(x_0) \, U_0(x_1). \notag
\end{align}


The linearized equations become:

\begin{equation}
    \begin{cases}
        \dfrac{\partial \boldsymbol{y}_{-1}}{\partial t} = \mu \dfrac{\partial^2}{\partial x_0^2} (\boldsymbol{y}_{0,0} + \boldsymbol{y}_{0,1}) + f(x_0) + h H(t) \delta_{x_0, x_1} \dfrac{\partial \boldsymbol{y}_1}{\partial x_1}, \\
        \dfrac{\partial \boldsymbol{y}_{0,0}}{\partial t} = \mu \dfrac{\partial^2 \boldsymbol{y}_{0,0}}{\partial x_0^2} + f(x_0), \\
        \dfrac{\partial \boldsymbol{y}_{0,1}}{\partial t} = \mu \dfrac{\partial^2 \boldsymbol{y}_{0,1}}{\partial x_0^2} + h H(t) \delta_{x_0, x_1} \dfrac{\partial \boldsymbol{y}_1}{\partial x_1}, \\
        \dfrac{\partial \boldsymbol{y}_1}{\partial t} = \mu \left( \dfrac{\partial^2}{\partial x_0^2} + \dfrac{\partial^2}{\partial x_1^2} \right) \boldsymbol{y}_1 + (\operatorP_{x_0,x_1}+1)f(x_1) \boldsymbol{y}_{0,0}.
    \end{cases}
\end{equation}

For clarity, we present these equations in matrix form:

\renewcommand{\arraystretch}{1.5}
\begin{equation}\label{eq-burgers-linear}
\frac{\partial }{\partial t}
\begin{bmatrix}
     \boldsymbol{y}_{-1}  \\
     \boldsymbol{y}_{0,0} \\
     \boldsymbol{y}_{0,1} \\
     \boldsymbol{y}_1
\end{bmatrix}
=
\begin{bmatrix}
     0 & \mu \dfrac{\partial^2}{\partial x_0^2} & \mu \dfrac{\partial^2}{\partial x_0^2} & h H(t) \delta_{x_0, x_1} \dfrac{\partial}{\partial x_1}  \\
     0 & \mu \dfrac{\partial^2}{\partial x_0^2} & 0 & 0 \\
     0 & 0 & \mu \dfrac{\partial^2}{\partial x_0^2} & h H(t) \delta_{x_0, x_1} \dfrac{\partial}{\partial x_1} \\
     0 & (\operatorP_{x_0,x_1}+1)f(x_1) & 0 & \mu \left( \dfrac{\partial^2}{\partial x_0^2} + \dfrac{\partial^2}{\partial x_1^2} \right)
\end{bmatrix}
\begin{bmatrix}
     \boldsymbol{y}_{-1}  \\
     \boldsymbol{y}_{0,0} \\
     \boldsymbol{y}_{0,1} \\
     \boldsymbol{y}_1
\end{bmatrix}
+
\begin{bmatrix}
     f(x_0)  \\
     f(x_0) \\
     0 \\
     0
\end{bmatrix}.
\end{equation}
\renewcommand{\arraystretch}{1}
Equation~(\ref{eq-burgers-linear}) represents a specific case of Eq.~(\ref{eq-linear-PDEs}). The corresponding relationships between the operators are defined as follows: \(\operatorN_{0,x_i} = f(x_i)\), \(\operatorN_{1,x_i} = \mu \partial^2_{x_i x_i}\), \(\operatorL_{1,1,x_i} = -\mathcal{I}\), and \(\operatorL_{2,1,x_j} = \partial_{x_j}\), where \(\mathcal{I}\) denotes the identity operator.
The initial and boundary conditions are derived from those of \( U_0 \) and \( U_1 \):

\renewcommand{\arraystretch}{1.5}
\begin{equation}
\boldsymbol{y}_{\text{in}} = 
\begin{bmatrix}
     0.3 \sin(\pi x_0)  \\
     0.3 \sin(\pi x_0) \\
     0 \\
     0.09 \sin(\pi x_0) \sin(\pi x_1)
\end{bmatrix}, \quad
\boldsymbol{y}(0, t) = \boldsymbol{y}(1, t) = 
\begin{bmatrix}
     0  \\
     0 \\
     0 \\
     0
\end{bmatrix}.
\end{equation}
\renewcommand{\arraystretch}{1}

At this stage, we have successfully linearized the forced Burgers' equation into a system of linear partial differential equations. This system can be solved using quantum linear PDE solvers within the QHAM framework, allowing us to obtain the solution \( u(x, t) \) efficiently on a quantum computer.

\subsubsection{Numerical tests}

Next, we test the performance of the QHAM in solving the Burgers' equation. In this test, we set \( H(t) = 1 \) and discretize \( x \) and \( t \) as follows:

\begin{align}
    x_i &= i \Delta x, \quad i = 1, 2, \ldots, n, \quad n = 32,\quad \Delta x = \frac{1}{n+1}, \notag\\
    t_j &= j \Delta t, \quad j = 0, 1, \ldots, T, \quad T = 100,\quad \Delta t = \frac{1}{T}.
\end{align}

The finite difference approximations for the spatial derivatives are:

\begin{align}
    \left. \frac{\partial u}{\partial x} \right|_{x = x_i} &\approx \frac{u(x_{i+1}) - u(x_{i-1})}{2\Delta x},\notag\\
    \left. \frac{\partial^2 u}{\partial x^2} \right|_{x = x_i} &\approx \frac{u(x_{i+1}) - 2u(x_i) + u(x_{i-1})}{\Delta x^2}.
\end{align}

\begin{figure}[h]
    \centering
    \includegraphics[width=16.9cm]{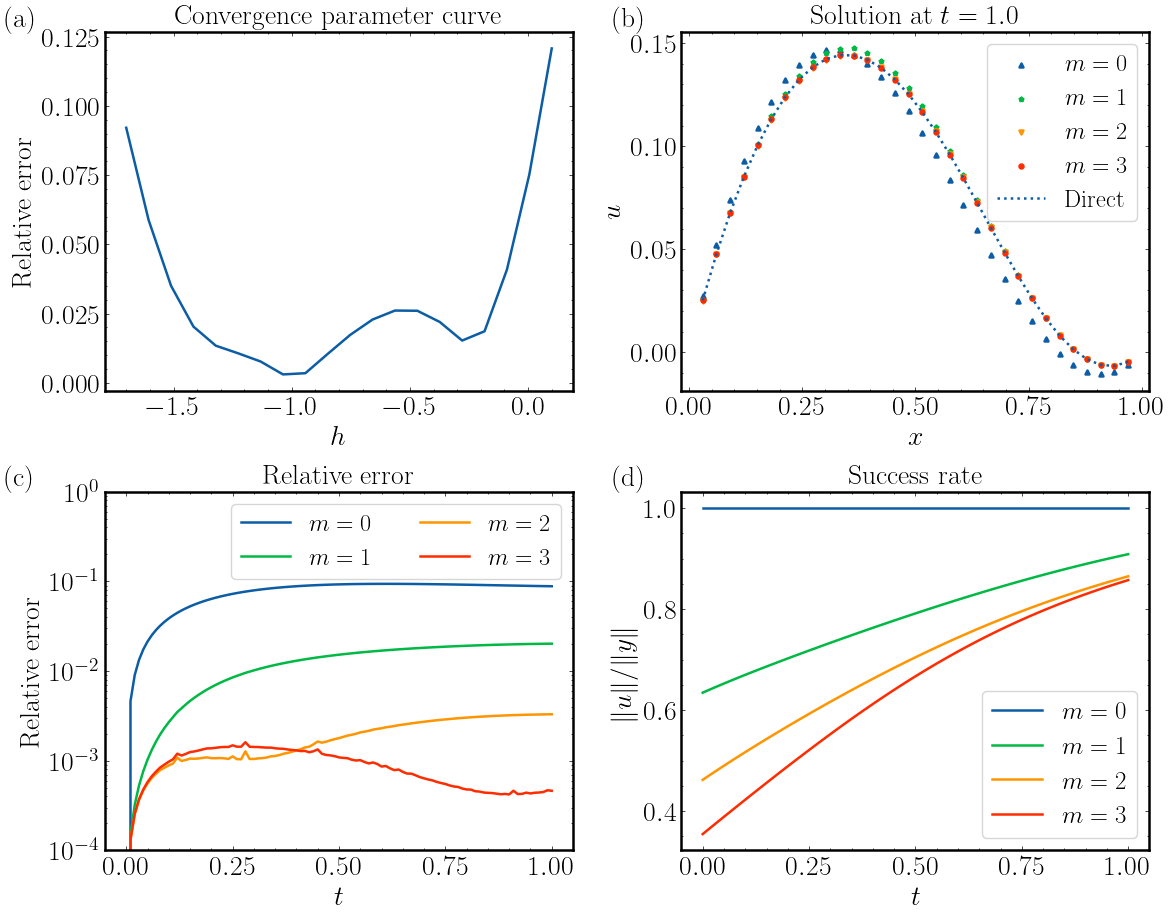}
    \caption{Performance of the QHAM for different truncation orders \( m \). (a) \( h \)-curve of the QHAM showing the relative error versus \( h \). (b) QHAM solution at \( t = 1.0 \) compared with the Runge-Kutta solution. (c) Relative error of the QHAM over time for different \( m \). (d) Success rate \( p \) of the QHAM over time for different \( m \).}
    \label{fig:burgers}
\end{figure}

The parameter \( h \) influences the convergence of the HAM. We first examine the \( h \)-curve, and the result is shown in Figure~\ref{fig:burgers}(a). The relative error, calculated over \( t \in [0, 1] \), is defined as:

\begin{equation}
    \text{Relative Error} = \frac{ \left[ \sum_{i=1}^{n} \sum_{j=0}^{T} \left( \sum_{k=0}^{m} U_k(x_i, t_j) - u(x_i, t_j) \right)^2 \right]^{1/2} }{ \left[ \sum_{i=1}^{n} \sum_{j=0}^{T} u^2(x_i, t_j) \right]^{1/2} }.
\end{equation}

We find that when \( h = -1 \), the relative error of the QHAM solution is minimized. Therefore, in subsequent tests, we set \( h = -1 \).

Figure~\ref{fig:burgers}(b) displays the QHAM solution at \( t = 1.0 \). We observe that when \( m = 3 \), the QHAM solution closely matches the solution obtained using the Runge-Kutta method. Figure~\ref{fig:burgers}(c) illustrates how the relative error changes over time for different values of \( m \); as \( m \) increases, the relative error decreases.

Finally, we evaluate the success rate \( p \) of the QHAM as \( m \) increases. Figure~\ref{fig:burgers}(d) shows the evolution of the success rate \( p \) over time \( t \) for various values of \( m \). Notably, \( m = 3 \) provides a sufficiently high success rate, which increases as \( t \) progresses.

The numerical results demonstrate that our proposed algorithm performs well in terms of both solution accuracy and success rate.

\subsection{KdV equations}\label{sec-kdv}

\subsubsection{Formulation}

The Korteweg–de Vries (KdV) equation is a fundamental nonlinear partial differential equation used to describe the propagation of solitary waves in shallow water and other contexts where weak nonlinearity and dispersion are present. The equation is expressed as:
\begin{equation}
    \frac{\partial u}{\partial t} + 6u \frac{\partial u}{\partial x} + \frac{\partial^3 u}{\partial x^3} = 0.
\end{equation}
The solitary wave solution to this equation is given by:
\begin{equation}
    u(x, t) = \frac{r}{2} \, \text{sech}^2\left[ \frac{\sqrt{r}}{2} \left( x - r t - \beta \right) \right],
\end{equation}
where \( r \) and \( \beta \) are constants. We set \( r = 0.5 \) and \( \beta = 0 \), so the initial condition becomes:
\begin{equation}
    u_{\text{in}}(x) = u(x, 0) = \frac{1}{4} \, \text{sech}^2\left( \frac{\sqrt{2}}{4} x \right).
\end{equation}

Next, we apply the quantum-compatible linearization of the IQHAM. We set the iteration number \( l = 1 \), the truncation order \( m = 1 \), and choose the linear operator \( \operatorL = \partial_t + \partial_{x}^3 \). The initial guess solution is:
\begin{equation}
    U_0(x, t) = u(x, 0).
\end{equation}
Following the process introduced in the application to the Burgers' equation, we define the variable \( \boldsymbol{y} \) of the original QHAM as:
\begin{align}
    \boldsymbol{y}_{-1} &= U_0(x_0) + U_1(x_0), \notag\\[6pt]
    \boldsymbol{y}_0 &= \begin{bmatrix} U_0(x_0) \\ U_1(x_0) \end{bmatrix}, \\[6pt]
    \boldsymbol{y}_1 &= U_0(x_0) \, U_0(x_1). \notag
\end{align}

The linearized equations become:
\begin{equation}\label{eq-kdv-linearPDE}
\left\{
\begin{aligned}
    \frac{\partial \boldsymbol{y}_{-1}}{\partial t} &= - \partial_{x_0}^3 \boldsymbol{y}_{0,1} + h H(t) \left[ \partial_{x_0}^3 \boldsymbol{y}_{0,0} + 6 \delta_{x_0, x_1} \partial_{x_1} \boldsymbol{y}_1 \right], \\
    \frac{\partial \boldsymbol{y}_{0,0}}{\partial t} &= 0, \\
    \frac{\partial \boldsymbol{y}_{0,1}}{\partial t} &= - \partial_{x_0}^3 \boldsymbol{y}_{0,1} + h H(t) \left[ \partial_{x_0}^3 \boldsymbol{y}_{0,0} + 6 \delta_{x_0, x_1} \partial_{x_1} \boldsymbol{y}_1 \right], \\
    \frac{\partial \boldsymbol{y}_1}{\partial t} &= 0.
\end{aligned}
\right.
\end{equation}

We can write Eq.~(\ref{eq-kdv-linearPDE}) more compactly as:
\begin{equation}
    \partial_t \boldsymbol{y} = \operatorL^{*}(\boldsymbol{y}),
\end{equation}
where \( \operatorL^{*} \) is the corresponding linear operator acting on \( \boldsymbol{y} \).

Next, we set \( \boldsymbol{y}_{-1} \) as the new initial guess solution \( \widetilde{U}_0 \) and define \( \widetilde{\boldsymbol{y}} \) as
\begin{align}
    \widetilde{\boldsymbol{y}}_{-1} &= \widetilde{U}_0(x_0) + \widetilde{U}_1(x_0), \notag\\
    \widetilde{\boldsymbol{y}}_0 &= \begin{bmatrix} \widetilde{U}_0(x_0) \\ \widetilde{U}_1(x_0) \end{bmatrix}, \\
    \widetilde{\boldsymbol{y}}_1 &= \widetilde{U}_0(x_0) \, \widetilde{U}_0(x_1). \notag
\end{align}
We then replace \( \widetilde{U}_0 \) in \( \widetilde{\boldsymbol{y}} \) with \( \boldsymbol{y} \) (except in \( \widetilde{\boldsymbol{y}}_{-1} \)) and obtain:
\begin{equation}
    \boldsymbol{z} = [\boldsymbol{z}_{-1}, \boldsymbol{z}_0, \boldsymbol{z}_1],
\end{equation}
where
\begin{align}
    \boldsymbol{z}_{-1} &= \widetilde{U}_0(x_0) + \widetilde{U}_1(x_0), \notag\\
    \boldsymbol{z}_0 &= \begin{bmatrix} \boldsymbol{y}(x_0, x_1) \\ \widetilde{U}_1(x_0) \end{bmatrix}, \quad \boldsymbol{z}_{0,0,i} = \boldsymbol{y}_{i}(x_0, x_1), \quad i = -1, 0, 1, \\
    \boldsymbol{z}_1 &= \boldsymbol{y}(x_0, x_1) \, \boldsymbol{y}(x_2, x_3), \quad \boldsymbol{z}_{1,i,j} = \boldsymbol{y}_{i}(x_0, x_1) \, \boldsymbol{y}_{j}(x_2, x_3), \quad i, j = -1, 0, 1. \notag
\end{align}

The variable \( \boldsymbol{z} \) satisfies the following linear PDEs:
\begin{equation}\label{eq-kdv-z}
\left\{
\begin{aligned}
    \frac{\partial \boldsymbol{z}_{-1}}{\partial t} &= - \partial_{x_0}^3 \boldsymbol{z}_{0,1} + [\operatorL^{*}(\boldsymbol{z}_{0,0})]_{-1} + h H(t) \operatorN(\boldsymbol{z}_{0,0,-1}), \\
    \frac{\partial \boldsymbol{z}_{0,0}}{\partial t} &= \operatorL^{*}(\boldsymbol{z}_{0,0}), \\
    \frac{\partial \boldsymbol{z}_{0,1}}{\partial t} &= - \partial_{x_0}^3 \boldsymbol{z}_{0,1} + h H(t) \operatorN(\boldsymbol{z}_{0,0,-1}), \\
    \frac{\partial \boldsymbol{z}_1}{\partial t} &= [\operatorL^{*}_{x_0, x_1} + \operatorL^{*}_{x_2, x_3}](\boldsymbol{z}_1).
\end{aligned}
\right.
\end{equation}
Here:
\begin{align}
    [\operatorL^{*}(\boldsymbol{z}_{0,0})]_{-1} &= \partial_t \widetilde{\boldsymbol{y}}_{-1}, \notag\\
    \operatorN(\boldsymbol{z}_{0,0,-1}) &= \left( \partial_t + \partial_{x_0}^3 \right) \boldsymbol{z}_{0,0,-1} + 6 \boldsymbol{z}_{0,0,-1} \, \partial_{x_0} \boldsymbol{z}_{0,0,-1} \notag\\
    &= \left( \partial_t + \partial_{x_0}^3 \right) \boldsymbol{z}_{0,0,-1} + 6 \delta_{x_0, x_2} \delta_{x_1, x_3} \partial_{x_0} \boldsymbol{z}_{1,-1,-1}.
\end{align}

Equation~(\ref{eq-kdv-z}) represents the linear PDEs for the first iteration (\( l = 1 \)) of the IQHAM. The initial conditions are:

\begin{align}
    \boldsymbol{z}_{-1,\text{in}} &= \boldsymbol{y}_{-1,\text{in}} = u_{\text{in}}(x_0), \notag\\
    \boldsymbol{z}_{0,0,\text{in}} &= \begin{bmatrix} u_{\text{in}}(x_0), u_{\text{in}}(x_0), 0, u_{\text{in}}(x_0) u_{\text{in}}(x_1) \end{bmatrix}, \notag\\
    \boldsymbol{z}_{0,1,\text{in}} &= 0, \notag\\
    \boldsymbol{z}_{1,\text{in}} &= \boldsymbol{z}_{0,0,\text{in}}(x_0, x_1) \, \boldsymbol{z}_{0,0,\text{in}}(x_2, x_3).
\end{align}

At this stage, we have successfully linearized the KdV equation into the system of linear PDEs defined in Eq.~(\ref{eq-kdv-z}).

\subsubsection{Numerical tests}

\begin{figure}[h]
    \centering
    \includegraphics[width=16.9cm]{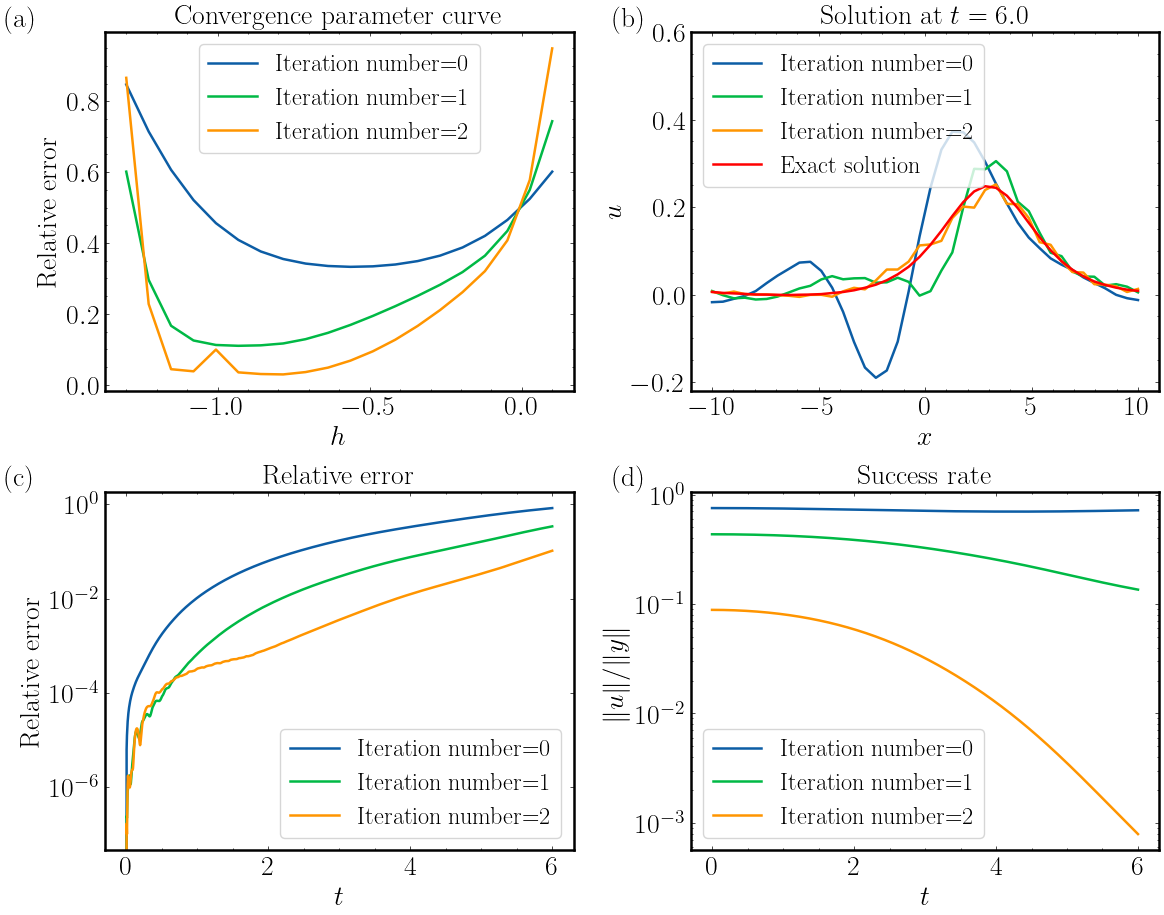}
    \caption{Performance of the IQHAM for different iteration numbers. (a) Convergence-control parameter \( h \)-curve. (b) IQHAM solution at \( t = 6.0 \). (c) Relative error of the IQHAM. (d) Success rate of the IQHAM.}
    \label{fig:kdv}
\end{figure}

Next, we evaluate the performance of the IQHAM through numerical simulations. In this test, we set \( H(t) = 1 \) and discretize the spatial domain as follows:
\[
x(i) = -10 + \Delta i, \quad i = 0, 1, 2, \dots, 40, \quad \Delta = 0.5.
\]
The finite difference approximations for the spatial derivatives are given by:
\[
\left. \frac{\partial u}{\partial x} \right|_{x = x(i)} \approx \frac{u(x(i+1)) - u(x(i-1))}{2\Delta},
\]
\[
\left. \frac{\partial^3 u}{\partial x^3} \right|_{x = x(i)} \approx \frac{u(x(i+2)) - 2u(x(i+1)) + 2u(x(i-1)) - u(x(i-2))}{2\Delta^3}.
\]
We set the evolution time interval to \( t \in [0, 6] \) with a time step of \( \Delta t = 0.005 \), the truncation order \( m = 3 \) and the iteration numbers \( l = 0, 1, 2 \). We first examine the convergence-control parameter \( h \) by plotting the \( h \)-curve, as shown in Figure~\ref{fig:kdv}(a). The \( h \)-curves for different iteration numbers exhibit similar behavior, and IQHAM performs well when \( h \in [-1, -0.5] \). Based on this observation, we set \( h = -0.8 \) for subsequent tests.

Figure~\ref{fig:kdv}(b) illustrates the IQHAM solution at time \( t = 6.0 \). As the iteration number \( l \) increases, the IQHAM solution progressively approaches the exact solution. Notably, with \( l = 2 \), we achieve a relatively accurate approximation. This convergence is further corroborated in Figure~\ref{fig:kdv}(c), where the relative error decreases as the iteration number increases.

We also assess the success rate of IQHAM, as shown in Figure~\ref{fig:kdv}(d). For iteration numbers \( l = 0 \) and \( l = 1 \), the success rate remains sufficiently high. However, when \( l = 2 \), the success rate diminishes as \( t \) increases. Therefore, although IQHAM offers an effective alternative to overcome issues with poor initial guess solutions, the computational cost and complexity escalate with higher iteration numbers. It is thus advisable to keep the iteration number as low as possible to balance accuracy and efficiency.

\section{Toward simulation of Navier-Stokes flow on a quantum computer}\label{sec:qham-NS}

The QHAM can be extended to Navier-Stokes flows. For simplicity, we consider the one-dimensional Navier-Stokes equations, which are written as
\begin{align}\label{1d-ns-eq}
\frac{\partial \mathbf{u}}{\partial t} + \frac{\partial \mathbf{F}}{\partial x} = 0,
\end{align}
where the conserved variable vector \( \mathbf{u} \) and the flux vector \( \mathbf{F} \) are defined as:
\begin{align}
\mathbf{u} =
\begin{bmatrix}
\rho \\
\rho u \\
\rho E \\
\end{bmatrix}, \quad
\mathbf{F} =
\begin{bmatrix}
\rho u \\
\rho u^2 + p - \tau_{xx} \\
\rho u H - \tau_{xx} u + q_x \\
\end{bmatrix}.
\end{align}
Here, \( \rho \), \( E \), and \( p \) represent the density, the specific total energy, and the pressure of the fluid, respectively. \( \tau_{xx} \), \( H \), and \( q_x \) represent the viscous stress, the specific total enthalpy, and the heat flux, respectively:
\begin{align}
&p = (\gamma - 1) \left( \rho E - \frac{1}{2} \rho u^2 \right), \notag\\
&H = \frac{1}{2} u^2 + \frac{p}{(\gamma - 1) \rho} + \frac{p}{\rho}, \notag\\
&\tau_{xx} = (2\mu + \lambda) \frac{\partial u}{\partial x}, \notag\\
&q_x = -k \frac{\partial T}{\partial x} = -\frac{\gamma \mu}{P_r (\gamma - 1)} \left( \frac{p}{\rho} \right)_x,
\end{align}
where \( \mu \), \( \lambda \), \( \gamma \), and \( P_r \) represent the dynamic (shear) viscosity coefficient, second (bulk) viscosity coefficient, ratio of specific heats, and Prandtl number, respectively.

Equation~(\ref{1d-ns-eq}) can then be written as
\begin{equation}
\operatorN(\mathbf{u}) = \left[
\begin{aligned}
\operatorN_1(\mathbf{u}) \\
\operatorN_2(\mathbf{u}) \\
\operatorN_3(\mathbf{u})
\end{aligned}
\right] = 0,
\end{equation}
where \( \operatorN_i(\mathbf{u}) \) is a polynomial of \( \mathbf{u} \) and its time/space derivatives. Next, we solve \( \operatorN(\mathbf{u}) = 0 \) using HAM and obtain the deformation equations related to \( U_i \). Our quantum-compatible linearization technique can also be used to linearize these \( U_i \)-related deformation equations. The difference from the process introduced in Section~\ref{sec-second-linearization} is that we need to add ancilla spaces for both \( x \) and \( t \), not just \( x \).

Specifically, we consider a cubic nonlinear component \( U_{k,0}(x, t) U_{l,0}(x, t) \frac{\partial}{\partial t} U_{s,1}(x, t) \). To linearize this term, we add ancilla spaces \( (x_1, t_1) \) and \( (x_2, t_2) \), and define
\begin{equation}
W(x, t, x_1, t_1, x_2, t_2) = U_{k,0}(x, t) U_{l,0}(x_1, t_1) U_{s,1}(x_2, t_2),
\end{equation}
which satisfies
\begin{equation}
U_{k,0} U_{l,0} \frac{\partial}{\partial t} U_{s,1} = \delta_{x, x_1} \delta_{x_1, x_2} \delta_{t, t_1} \delta_{t_1, t_2} \frac{\partial}{\partial t_2} W.
\end{equation}
Here, \( \delta_{x, x_1} \delta_{x_1, x_2} \delta_{t, t_1} \delta_{t_1, t_2} \frac{\partial}{\partial t_2} \) is a linear operator; therefore, \( U_{k,0} U_{l,0} \frac{\partial}{\partial t} U_{s,1} \) is linear in \( W \). We use the new variable construction process introduced earlier and complete the quantum-compatible linearization process as shown in Figure~\ref{fig:secondary-linearization}. The upper triangular structure of the deformation equations also ensures that the quantum-compatible linearization will terminate, thus achieving the linearization of the deformation equations.

Therefore, the one-dimensional Navier-Stokes equations can be linearized into linear PDEs, after which we can use quantum linear PDE solvers to obtain the desired solution. The two-dimensional and three-dimensional Navier-Stokes equations can also be linearized into linear PDEs using a similar approach.

However, since the solution of the Navier-Stokes equations is more complex, choosing appropriate convergence parameters, initial guess solutions, and other HAM parameters to ensure convergence is challenging. Using the current QHAM to solve the Navier-Stokes equations may face problems of non-convergence and low success rates. In the future, we will introduce improvements to HAM within QHAM to optimize its performance and complete the task of solving the Navier-Stokes equations.

\section{Conclusion and discussion}\label{sec:conclusion}

In this study, we introduced a Quantum Homotopy Analysis Method (QHAM) with quantum-compatible linearization to tackle the challenge of solving nonlinear PDEs using quantum computing. QHAM integrates the Homotopy Analysis Method (HAM) with a novel quantum-compatible linearization process, transforming nonlinear PDEs into a system of linear PDEs that can be efficiently solved using quantum algorithms. By analyzing the computational complexity, we demonstrated that QHAM preserves the exponential speedup offered by quantum linear PDE solvers while ensuring that the computational complexity increases only polynomially with the HAM truncation order.

The efficiency of QHAM depends on an optimal initial guess, which is generally difficult to obtain. To address this, we proposed the Iterative Quantum Homotopy Analysis Method (IQHAM), which iteratively refines the approximate solution by using the result of the previous QHAM iteration as the new initial guess. Although IQHAM requires additional computational resources with each iteration, it offers improved convergence and reduces the dependence on an accurate initial guess. This approach provides a trade-off between accuracy, prior knowledge requirements, and computational complexity, achieving a balanced method for tackling challenging nonlinear problems.

Building upon our theoretical framework, we applied QHAM to the Burgers' equation and the KdV equation, demonstrating its convergence, accuracy, and success rate. These practical implementations highlight the influence of HAM parameters on solution quality and confirm QHAM's potential for accurately solving nonlinear PDEs using quantum computers. Furthermore, we presented a technical approach to extend the QHAM to solve the Navier-Stokes equations, showcasing the method's scalability to more complex and higher-dimensional nonlinear PDEs.

The core advancement of QHAM is the quantum-compatible linearization strategy, which embeds all computational tasks of the HAM into a system of linear PDEs by introducing auxiliary spaces. This quantum-compatible linearization effectively transforms the iterative execution of local linearizations into a quantum-compatible linearization process, thereby avoiding the exponential increase in complexity that arises from directly accelerating each local linearization step using quantum computing. This technique can be applied to other methods, such as embedding explicit time discretization methods for solving nonlinear differential equations or transforming the Newton method optimization process into a linear system. An open question remains regarding how to construct the variables during the quantum-compatible linearization process to maximize the probability of obtaining the target solution from the linearized PDEs.

In conclusion, QHAM provides a promising pathway for leveraging quantum computing to solve nonlinear PDEs, particularly in the field of computational fluid dynamics. Future work will focus on enhancing the QHAM framework by optimizing HAM parameters and integrating advanced quantum solvers, aiming to extend the method's applicability to a broader range of nonlinear problems, including multi-dimensional Navier-Stokes equations. As quantum computing technology continues to advance, we anticipate that QHAM and related methods will become viable tools for addressing some of the most challenging problems in computational science.

\section*{Acknowledgments}
This work has been supported by the National Key Research and Development Program of China (Grant No. 2023YFB4502500), the National Natural Science Foundation of China (Grant No. 12404564), and the Anhui Province Science and Technology Innovation (Grant No. 202423s06050001).

\appendix

\bibliography{qham}

@article{witherden2014pyfr,
  title={{PyFR: An open source framework for solving advection--diffusion type problems on streaming architectures using the flux reconstruction approach}},
  author={Witherden, Freddie D and Farrington, Antony M and Vincent, Peter E},
  journal={Computer Physics Communications},
  volume={185},
  number={11},
  pages={3028--3040},
  year={2014},
  publisher={Elsevier}
}

@techreport{slotnick2014cfd,
  title = {{{CFD Vision}} 2030 {{Study}}: {{A Path}} to {{Revolutionary Computational Aerosciences}}},
  author = {Slotnick, Jeffrey and Khodadoust, Abdollah and Alonso, Juan and Darmofal, David and Gropp, William and Lurie, Elizabeth and Mavriplis, Dimitri},
  year = {2014},
  number = {CR\textendash 2014-218178},
  pages = {58},
  address = {{Langley Research Center, Hampton, VA}},
  institution = {{National Aeronautics and Space Administration (NASA)}}
}

@article{chen2024enabling,
title = {{Enabling large-scale and high-precision fluid simulations on near-term quantum computers}},
journal = {Computer Methods in Applied Mechanics and Engineering},
volume = {432},
pages = {117428},
year = {2024},
author = {Zhao-Yun Chen and Teng-Yang Ma and Chuang-Chao Ye and Liang Xu and Wen Bai and Lei Zhou and Ming-Yang Tan and Xi-Ning Zhuang and Xiao-Fan Xu and Yun-Jie Wang and Tai-Ping Sun and Yong Chen and Lei Du and Liang-Liang Guo and Hai-Feng Zhang and Hao-Ran Tao and Tian-Le Wang and Xiao-Yan Yang and Ze-An Zhao and Peng Wang and Sheng Zhang and Ren-Ze Zhao and Chi Zhang and Zhi-Long Jia and Wei-Cheng Kong and Meng-Han Dou and Jun-Chao Wang and Huan-Yu Liu and Cheng Xue and Peng-Jun-Yi Zhang and Sheng-Hong Huang and Peng Duan and Yu-Chun Wu and Guo-Ping Guo},
}

@article{wu2022vertical,
  title={Vertical {MoS2} transistors with sub-1-nm gate lengths},
  author={Wu, Fan and Tian, He and Shen, Yang and Hou, Zhan and Ren, Jie and Gou, Guangyang and Sun, Yabin and Yang, Yi and Ren, Tian-Ling},
  journal={Nature},
  volume={603},
  number={7900},
  pages={259--264},
  year={2022},
  publisher={Nature Publishing Group UK London}
}

@article{shor1999polynomial,
  title={Polynomial-time algorithms for prime factorization and discrete logarithms on a quantum computer},
  author={Shor, Peter W},
  journal={SIAM review},
  volume={41},
  number={2},
  pages={303--332},
  year={1999},
  publisher={SIAM}
}

@article{meng2023quantum,
  title={Quantum computing of fluid dynamics using the hydrodynamic Schr{\"o}dinger equation},
  author={Meng, Zhaoyuan and Yang, Yue},
  journal={Physical Review Research},
  volume={5},
  number={3},
  pages={033182},
  year={2023},
  publisher={APS}
}

@article{giannakis2022embedding,
  title={Embedding classical dynamics in a quantum computer},
  author={Giannakis, Dimitrios and Ourmazd, Abbas and Pfeffer, Philipp and Schumacher, J{\"o}rg and Slawinska, Joanna},
  journal={Physical Review A},
  volume={105},
  number={5},
  pages={052404},
  year={2022},
  publisher={APS}
}

@article{meng2024simulating,
  title={Simulating unsteady flows on a superconducting quantum processor},
  author={Meng, Zhaoyuan and Zhong, Jiarun and Xu, Shibo and Wang, Ke and Chen, Jiachen and Jin, Feitong and Zhu, Xuhao and Gao, Yu and Wu, Yaozu and Zhang, Chuanyu and others},
  journal={Communications Physics},
  volume={7},
  number={1},
  pages={349},
  year={2024},
  publisher={Nature Publishing Group UK London}
}

@article{cao2013quantum,
  title={Quantum algorithm and circuit design solving the Poisson equation},
  author={Cao, Yudong and Papageorgiou, Anargyros and Petras, Iasonas and Traub, Joseph and Kais, Sabre},
  journal={New Journal of Physics},
  volume={15},
  number={1},
  pages={013021},
  year={2013},
  publisher={IOP Publishing}
}

@article{childs2020quantum,
  title={Quantum spectral methods for differential equations},
  author={Childs, Andrew M and Liu, Jin-Peng},
  journal={Communications in Mathematical Physics},
  volume={375},
  number={2},
  pages={1427--1457},
  year={2020},
  publisher={Springer}
}

@article{harrow2009quantum,
  title={Quantum algorithm for linear systems of equations},
  author={Harrow, Aram W and Hassidim, Avinatan and Lloyd, Seth},
  journal={Physical review letters},
  volume={103},
  number={15},
  pages={150502},
  year={2009},
  publisher={APS}
}

@article{an2022quantum,
  title={Quantum linear system solver based on time-optimal adiabatic quantum computing and quantum approximate optimization algorithm},
  author={An, Dong and Lin, Lin},
  journal={ACM Transactions on Quantum Computing},
  volume={3},
  number={2},
  pages={1--28},
  year={2022},
  publisher={ACM New York, NY}
}

@article{costa2022optimal,
  title={Optimal scaling quantum linear-systems solver via discrete adiabatic theorem},
  author={Costa, Pedro CS and An, Dong and Sanders, Yuval R and Su, Yuan and Babbush, Ryan and Berry, Dominic W},
  journal={PRX quantum},
  volume={3},
  number={4},
  pages={040303},
  year={2022},
  publisher={APS}
}

@article{jin2022quantum,
  title={Quantum simulation of partial differential equations via schr{\"o}dingerisation: technical details},
  author={Jin, Shi and Liu, Nana and Yu, Yue},
  journal={arXiv preprint arXiv:2212.14703},
  year={2022}
}

@article{jin2023quantum,
  title={Quantum simulation of partial differential equations: Applications and detailed analysis},
  author={Jin, Shi and Liu, Nana and Yu, Yue},
  journal={Physical Review A},
  volume={108},
  number={3},
  pages={032603},
  year={2023},
  publisher={APS}
}

@article{jin2024schr,
  title={On Schr{\"o}dingerization based quantum algorithms for linear dynamical systems with inhomogeneous terms},
  author={Jin, Shi and Liu, Nana and Ma, Chuwen},
  journal={arXiv preprint arXiv:2402.14696},
  year={2024}
}

@article{an2023linear,
  title={Linear combination of Hamiltonian simulation for nonunitary dynamics with optimal state preparation cost},
  author={An, Dong and Liu, Jin-Peng and Lin, Lin},
  journal={Physical Review Letters},
  volume={131},
  number={15},
  pages={150603},
  year={2023},
  publisher={APS}
}

@article{chen2022quantum,
  title={Quantum approach to accelerate finite volume method on steady computational fluid dynamics problems},
  author={Chen, Zhao-Yun and Xue, Cheng and Chen, Si-Ming and Lu, Bing-Han and Wu, Yu-Chun and Ding, Ju-Chun and Huang, Sheng-Hong and Guo, Guo-Ping},
  journal={Quantum Information Processing},
  volume={21},
  number={4},
  pages={137},
  year={2022},
  publisher={Springer}
}

@article{joczik2022cost,
  title={A cost-efficient approach towards computational fluid dynamics simulations on quantum devices},
  author={J{\'o}czik, Szabolcs and Zimbor{\'a}s, Zolt{\'a}n and Majoros, Tam{\'a}s and Kiss, Attila},
  journal={Applied Sciences},
  volume={12},
  number={6},
  pages={2873},
  year={2022},
  publisher={MDPI}
}

@article{liu2021efficient,
  title={Efficient quantum algorithm for dissipative nonlinear differential equations},
  author={Liu, Jin-Peng and Kolden, Herman {\O}ie and Krovi, Hari K and Loureiro, Nuno F and Trivisa, Konstantina and Childs, Andrew M},
  journal={Proceedings of the National Academy of Sciences},
  volume={118},
  number={35},
  pages={e2026805118},
  year={2021},
  publisher={National Acad Sciences}
}

@article{joseph2020koopman,
  title={Koopman--von Neumann approach to quantum simulation of nonlinear classical dynamics},
  author={Joseph, Ilon},
  journal={Physical Review Research},
  volume={2},
  number={4},
  pages={043102},
  year={2020},
  publisher={APS}
}

@article{engel2021linear,
    author = {Engel, Alexander and Smith, Graeme and Parker, Scott E.},
    title = {Linear embedding of nonlinear dynamical systems and prospects for efficient quantum algorithms},
    journal = {Physics of Plasmas},
    volume = {28},
    number = {6},
    pages = {062305},
    year = {2021},
    month = {06}
}

@article{jin2024quantum,
  title={Quantum algorithms for nonlinear partial differential equations},
  author={Jin, Shi and Liu, Nana},
  journal={Bulletin des Sciences Math{\'e}matiques},
  volume={194},
  pages={103457},
  year={2024},
  publisher={Elsevier}
}

@article{xue2021quantum,
  title={Quantum homotopy perturbation method for nonlinear dissipative ordinary differential equations},
  author={Xue, Cheng and Wu, Yu-Chun and Guo, Guo-Ping},
  journal={New Journal of Physics},
  volume={23},
  number={12},
  pages={123035},
  year={2021},
  publisher={IOP Publishing}
}

@article{liao2024general,
  title={A general frame of quantum simulation for nonlinear partial differential equations},
  author={Liao, Shijun},
  journal={arXiv preprint arXiv:2406.15821},
  year={2024}
}

@phdthesis{liao1992proposed,
  title={The proposed homotopy analysis technique for the solution of nonlinear problems},
  author={Liao, Shijun},
  year={1992},
  school={Ph. D. Thesis, Shanghai Jiao Tong University Shanghai}
}

@book{liao2003beyond,
  title={Beyond Perturbation: Introduction to the Homotopy Analysis Method},
  author={Liao, Shijun},
  year={2003},
  edition={1st},
  publisher={Chapman and Hall/CRC},
  address={New York},
  pages={336},
  isbn={9780429208614}
}

@article{liao2004homotopy,
  title={On the homotopy analysis method for nonlinear problems},
  author={Liao, Shijun},
  journal={Applied mathematics and computation},
  volume={147},
  number={2},
  pages={499--513},
  year={2004},
  publisher={Elsevier}
}

@article{steijl2018parallel,
  title={Parallel evaluation of quantum algorithms for computational fluid dynamics},
  author={Steijl, Ren{\'e} and Barakos, George N},
  journal={Computers \& Fluids},
  volume={173},
  pages={22--28},
  year={2018},
  publisher={Elsevier}
}

@article{budinski2021quantum,
  title={Quantum algorithm for the advection--diffusion equation simulated with the lattice Boltzmann method},
  author={Budinski, Ljubomir},
  journal={Quantum Information Processing},
  volume={20},
  number={2},
  pages={57},
  year={2021},
  publisher={Springer}
}

@article{gaitan2021finding,
  title={Finding Solutions of the Navier-Stokes Equations through Quantum Computing—Recent Progress, a Generalization, and Next Steps Forward},
  author={Gaitan, Frank},
  journal={Advanced Quantum Technologies},
  volume={4},
  number={10},
  pages={2100055},
  year={2021},
  publisher={Wiley Online Library}
}

@article{jin2024analog,
  title={Analog quantum simulation of partial differential equations},
  author={Shi Jin and Nana Liu},
  journal={Quantum Science and Technology},
  year = {2024},
  month = {jun},
  publisher = {IOP Publishing},
  volume = {9},
  number = {3},
  pages = {035047}
}

@article{sarma2024quantum,
  title={Quantum variational solving of nonlinear and multidimensional partial differential equations},
  author={Sarma, Abhijat and Watts, Thomas W and Moosa, Mudassir and Liu, Yilian and McMahon, Peter L},
  journal={Physical Review A},
  volume={109},
  number={6},
  pages={062616},
  year={2024},
  publisher={APS}
}

@article{lubasch2020variational,
  title={Variational quantum algorithms for nonlinear problems},
  author={Lubasch, Michael and Joo, Jaewoo and Moinier, Pierre and Kiffner, Martin and Jaksch, Dieter},
  journal={Physical Review A},
  volume={101},
  number={1},
  pages={010301},
  year={2020},
  publisher={APS}
}

@article{kyriienko2021solving,
  title={Solving nonlinear differential equations with differentiable quantum circuits},
  author={Kyriienko, Oleksandr and Paine, Annie E and Elfving, Vincent E},
  journal={Physical Review A},
  volume={103},
  number={5},
  pages={052416},
  year={2021},
  publisher={APS}
}

@article{leyton2008quantum,
  title={A quantum algorithm to solve nonlinear differential equations},
  author={Leyton, Sarah K and Osborne, Tobias J},
  journal={arXiv preprint arXiv:0812.4423},
  year={2008}
}

@article{oz2023efficient,
  title={An efficient quantum partial differential equation solver with chebyshev points},
  author={Oz, Furkan and San, Omer and Kara, Kursat},
  journal={Scientific Reports},
  volume={13},
  number={1},
  pages={7767},
  year={2023},
  publisher={Nature Publishing Group UK London}
}

@article{jin2212quantum,
  title={Quantum simulation of partial differential equations via schr{\"o}dingerisation (2022)},
  author={Jin, S and Liu, N and Yu, Y},
  journal={arXiv preprint arXiv:2212.13969}
}

@article{berry2024quantum,
  title={Quantum algorithm for time-dependent differential equations using Dyson series},
  author={Berry, Dominic W and Costa, Pedro CS},
  journal={Quantum},
  volume={8},
  pages={1369},
  year={2024},
  publisher={Verein zur F{\"o}rderung des Open Access Publizierens in den Quantenwissenschaften}
}

@article{an2023quantum,
  title={Quantum algorithm for linear non-unitary dynamics with near-optimal dependence on all parameters},
  author={An, Dong and Childs, Andrew M and Lin, Lin},
  journal={arXiv preprint arXiv:2312.03916},
  year={2023}
}

@article{shang2024design,
  title={Design nearly optimal quantum algorithm for linear differential equations via Lindbladians},
  author={Shang, Zhong-Xia and Guo, Naixu and An, Dong and Zhao, Qi},
  journal={arXiv preprint arXiv:2410.19628},
  year={2024}
}

@article{linden2022quantum,
  title={Quantum vs. classical algorithms for solving the heat equation},
  author={Linden, Noah and Montanaro, Ashley and Shao, Changpeng},
  journal={Communications in Mathematical Physics},
  volume={395},
  number={2},
  pages={601--641},
  year={2022},
  publisher={Springer}
}

@article{wang2020quantum,
  title={Quantum fast Poisson solver: the algorithm and complete and modular circuit design},
  author={Wang, Shengbin and Wang, Zhimin and Li, Wendong and Fan, Lixin and Wei, Zhiqiang and Gu, Yongjian},
  journal={Quantum Information Processing},
  volume={19},
  pages={1--25},
  year={2020},
  publisher={Springer}
}

@article{liu2021variational,
  title={Variational quantum algorithm for the Poisson equation},
  author={Liu, Hai-Ling and Wu, Yu-Sen and Wan, Lin-Chun and Pan, Shi-Jie and Qin, Su-Juan and Gao, Fei and Wen, Qiao-Yan},
  journal={Physical Review A},
  volume={104},
  number={2},
  pages={022418},
  year={2021},
  publisher={APS}
}

@book{thomas1995numerical,
  title={Numerical Partial Differential Equations: Finite Difference Methods},
  author={Thomas, J. W.},
  volume={22},
  year={1995},
  publisher={Springer New York},
  address={New York},
  isbn={978-0-387-97999-1},
  edition={1},
  pages={437}
}

@article{chakraborty2018power,
  title={The power of block-encoded matrix powers: improved regression techniques via faster Hamiltonian simulation},
  author={Chakraborty, Shantanav and Gily{\'e}n, Andr{\'a}s and Jeffery, Stacey},
  journal={arXiv preprint arXiv:1804.01973},
  year={2018}
}

@article{grover2002creating,
  title={Creating superpositions that correspond to efficiently integrable probability distributions},
  author={Grover, Lov and Rudolph, Terry},
  journal={arXiv preprint quant-ph/0208112},
  year={2002}
}

@article{liao2009notes,
  title={Notes on the homotopy analysis method: some definitions and theorems},
  author={Liao, Shijun},
  journal={Communications in Nonlinear Science and Numerical Simulation},
  volume={14},
  number={4},
  pages={983--997},
  year={2009},
  publisher={Elsevier}
}

@article{liao2020new,
  title={A new non-perturbative approach in quantum mechanics for time-independent Schr{\"o}dinger equations},
  author={Liao, ShiJun},
  journal={Science China Physics, Mechanics \& Astronomy},
  volume={63},
  number={3},
  pages={234612},
  year={2020},
  publisher={Springer}
}

@article{huang2023near,
  title={Near-term quantum computing techniques: Variational quantum algorithms, error mitigation, circuit compilation, benchmarking and classical simulation},
  author={Huang, He-Liang and Xu, Xiao-Yue and Guo, Chu and Tian, Guojing and Wei, Shi-Jie and Sun, Xiaoming and Bao, Wan-Su and Long, Gui-Lu},
  journal={Science China Physics, Mechanics \& Astronomy},
  volume={66},
  number={5},
  pages={250302},
  year={2023},
  publisher={Springer}
}

@inproceedings{gilyen2019quantum,
    author = {Gily\'{e}n, Andr\'{a}s and Su, Yuan and Low, Guang Hao and Wiebe, Nathan},
    title = {Quantum singular value transformation and beyond: exponential improvements for quantum matrix arithmetics},
    year = {2019},
    isbn = {9781450367059},
    publisher = {Association for Computing Machinery},
    address = {New York, NY, USA},
    url = {https://doi.org/10.1145/3313276.3316366},
    doi = {10.1145/3313276.3316366},
    booktitle = {Proceedings of the 51st Annual ACM SIGACT Symposium on Theory of Computing},
    pages = {193–204},
    numpages = {12},
    location = {Phoenix, AZ, USA},
    series = {STOC 2019}
}

\end{document}